\documentclass[a4paper,english]{article}
\usepackage[LGR,T1]{fontenc}
\usepackage[latin9]{inputenc}
\usepackage{xcolor}
\usepackage{array}
\usepackage{float}
\usepackage{calc}
\usepackage{textcomp}
\usepackage{multirow}
\usepackage{tipa}
\usepackage{amsbsy}
\usepackage{amstext}
\usepackage{graphicx}

\makeatletter

\pdfpageheight\paperheight
\pdfpagewidth\paperwidth

\DeclareRobustCommand{\greektext}{%
  \fontencoding{LGR}\selectfont\def\encodingdefault{LGR}}
\DeclareRobustCommand{\textgreek}[1]{\leavevmode{\greektext #1}}
\ProvideTextCommand{\~}{LGR}[1]{\char126#1}

\newcommand{\lyxmathsym}[1]{\ifmmode\begingroup\def\b@ld{bold}
  \text{\ifx\math@version\b@ld\bfseries\fi#1}\endgroup\else#1\fi}

\providecommand{\tabularnewline}{\\}

\@ifundefined{showcaptionsetup}{}{%
 \PassOptionsToPackage{caption=false}{subfig}}
\usepackage{subfig}
\makeatother

\usepackage{babel}
\begin{document}

\title{\bigskip{}
A model of infant speech perception and learning{\large{}}\linebreak{}
\bigskip{}
\bigskip{}
}

\author{Philip Zurbuchen\bigskip{}
}

\date{September 16, 2016}

\maketitle
 \thispagestyle{empty}\clearpage{}

\part*{Acknowledgements}

\bigskip{}

I'd like to thank Jochen Triesch for his supervision over the last
year, during which he introduced me to the field of computational
neuroscience. I thoroughly enjoyed his lectures (for instance on Reinforcement
Learning) and his insights and advice whenever we discussed my work.

Many thanks to Max Murakami. Having (in his master thesis) developed
the original speech acquisition model, Max did a great job of patiently
lecturing me on the ins and outs of \textit{Listen and Babble} and
its context: speech acquisition in the human brain. I learned a lot
this past year in many areas; much of that is attributed to him.

I thank my mother, Sarah Zurbuchen, for turning Swiss-English into
proper English over the years, but also for correcting remaining spelling
errors in this thesis.

Special thanks to my dear wife, Tirza, who was willing to put up with
my long programming evenings and politely listened to my musings about
how 'this-and-that' might be connected to 'that-and-that' (before
changing my mind again). She is a social worker and I'm glad she can
cope with a case like me!

Lastly, as did greater men like Newton, Maxwell and Faraday, I dedicate
my efforts to the Creator of speech, who himself spoke through Christ
Jesus.

\date{ \thispagestyle{empty}}

\clearpage{}

\tableofcontents{}\frenchspacing

\date{ \thispagestyle{empty}}

\clearpage{}

\date{ \setcounter{page}{1} }

\section{Introduction}

\textcompwordmark{}

\subsection{Infants learn speech}

You learned to speak as an infant. We all did \textendash{} we learned
speech without knowledge or concept of the very language in which
we now so naturally articulate our thoughts, feelings and questions.
This in itself is one of the many amazing facts about this universe.

In my thesis, I introduce the reader to the learning process involved
in acquiring speech. We'll focus on specific challenges involved in
learning to speak a language from scratch. And while we're at it,
I'll point out a few important aspects of speech itself. How is speech
produced in a human being, and what is man's instrument when he says
``I love a good conversation!''~?

Then in section~\ref{subsec:Model_Explained}, we'll review previous
work done by M. Murakami in 2014: the original \textit{Listen and
Babble} model. My work is complementary to \textit{Listen and Babble}
in that I seek to discuss and extend the model. Restating all the
details of Murakami's work would go too far \textendash{} I refer
to his thesis and publication for more detail~\cite{Paper Murakami,Thesis Murakami}.
I shall cover some important points already mentioned in Murakami's
work, specifically concerning the shortcomings of the model and ideas
on improving parts of the model. My research addresses some of these
shortcomings and seeks to improve the conceptual framework of \textit{Listen
and Babble} by proposing new ways of thinking about such issues.

But first, we must start at the beginning and ask the question: when
do infants start learning to speak a language?

\subsubsection{\label{subsec:The-challenge-of}The challenge of speech perception}

We started to learn speech very early, even before our first utterances.
Research shows that infants start mapping critical aspects of ambient
speech in the first year of life~\cite{P.Kuhl Language/Culture/Brain}.
Ambient language sounds are listened to and analysed before ever understanding
a word~\cite{P.Kuhl_other_1,P.Kuhl_other_2}. 

Infants are confronted with lots of challenges when trying to hear
and recognise speech sounds. Of course the infant must learn to perceive
speech well in order to start loading syllables and words with meaning,
or even imitating the caregiver's speech.

\paragraph{Recognising speech sounds}

We are quick to think that infants hear well defined and clear language
signals. We think they ``hear what we say'' and merely have to learn
the meaning of the words \textendash{} perhaps a bit like a code breaker,
trying to make sense of a jumble of (seemingly randomly arranged)
typed out letters. But actually the task of hearing language is much
harder. Infants have to learn speech \textit{perception} first.

This becomes more clear when we are presented with examples from adult
foreign-language perception. Native German speakers have problems
hearing the difference between English speech sounds {[}ð{]}\footnote{Phones/Phonemes given in IPA (International Phonetic Alphabet) notation.
See appendix for an overview of all IPA symbols.} (as in \textbf{th}is) and {[}\textgreek{j}{]} (\textbf{th}ing). While
/ð/ is voiced, /\textgreek{j}/ is voiceless. Nonetheless, these both
sound like the same speech sound for many non-English speakers\footnote{In fact, even English-speaking children have difficulties in distinguishing
between /ð/ and /\textgreek{j}/ (which are among the last phonemes
learned in the English language). These are often confused by young
children \textendash{} frequently even until they start elementary
school.

``Prior to this age {[}5 years{]}, many children substitute the sounds
{[}f{]} and {[}v{]} respectively. For small children, \textit{fought}
and \textit{thought} are therefore homophones {[}phones which are
perceived as the same phoneme{]}. As British and American children
begin school at age four and five respectively, this means that many
are learning to read and write before they have sorted out these sounds,
and the infantile pronunciation is frequently reflected in their spelling
errors: \textit{ve fing} for \textit{the thing}.''~\cite{WIKI_th}}.

People who grow up with different ambient language perceive speech
differently. An infant must learn perception of his ambient language
as a first step. It is easy to see that the example of the code breaker,
who reads well defined, typed out alphabetical symbols, does not go
far enough. Instead, the infant code breaker is confronted with a
continuous signal, without any gaps between letters (or words!).

When listening to our native language, we think we hear separate words
when spoken. But this is not the case. I recorded myself saying ``I
like apple pie'' and plotted waveform and spectrogram (see Box~1).
Confronted with the physical signal, finding the end of one word and
the beginning of the next is not trivial. Also, notice that we all
would hear two p-sounds in the sentence ``I like a\textbf{pp}le \textbf{p}ie''.
But finding the p-sounds in the spectrogram is not that easy. And
when we've found them, we realise that they aren't the same at all,
due to their \textit{phonetic context }(i.e. which sounds come before
or after).
\begin{center}
\begin{figure}
\centering{}%
\fbox{\begin{minipage}[t]{0.95\columnwidth}%
\textbf{Waveform and Spectrogram\hfill{}Box~1\medskip{}
}

We can analyse sound using either its waveform or spectrogram. A waveform
is simply the pressure wave of the speech sound (which arrived at
the microphone/human ear) plotted over time. The waveform can be broken
down into all the single frequencies that contribute to the waveform.
The power in each of these frequencies (\textit{spectral density})
can be plotted over time, too. A plot of the spectral density over
time is called a spectrogram.

We do well to look at spectrograms when thinking about speech input.
Sound, after arriving in form of mechanical waves in the inner ear,
is transformed in the cochlea into neural spiking. Each neuron location
in the cochlea responds to a different frequency. This tonotopic organisation
(spatial layout of frequencies) is repeated in other areas of the
brain (inferior colliculus, primary auditory cortex). The brain effectively
works with something like a spectrogram of the sound input. Spectral
density over time is therefore a suitable representation of speech
signals.

Frequencies at spectral power maxima (darkest colouring in the spectrogram
below) are called formant frequencies ($F_{1}\,,F_{2,}\,F_{3}$).
The maximum which is lowest in frequency is called the pitch, or $F_{0}$.
We can think of the pitch as a measure of ``how high a sound is''.
If I sing the musical note $C_{1}$ for example, I'll have an $F_{0}$
of 32.70~Hz.

\subfloat[\label{fig:Applepie}Waveform (above) and spectrogram (below) of the
sentence ``I like apple pie''. In the spectrogram, darker colour
denotes higher spectral density. Word borders are marked by vertical
dashed lines. Extracted and drawn using Praat~\cite{Praat}.]{\centering{}\includegraphics[width=1\columnwidth]{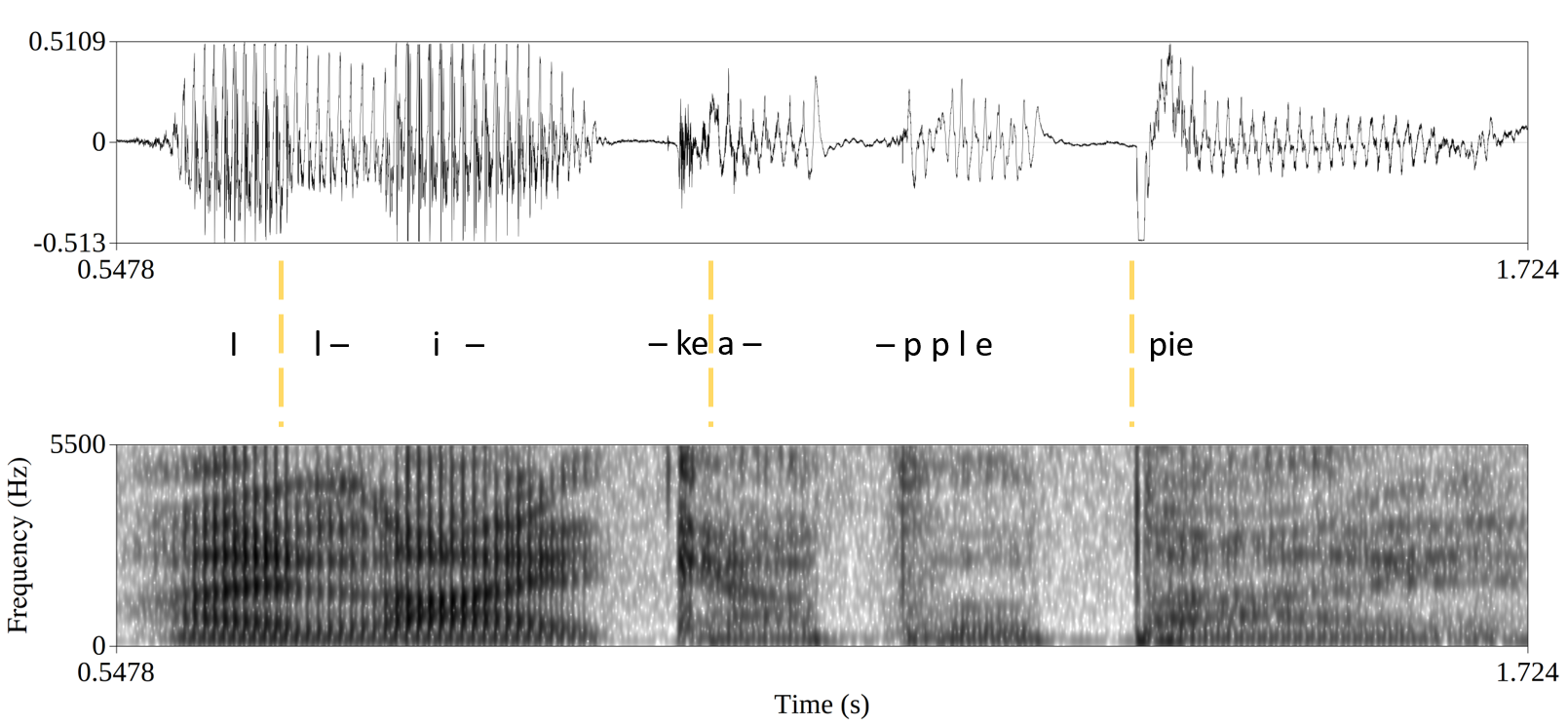}}%
\end{minipage}}
\end{figure}
\par\end{center}

\paragraph{Phones and phonemes}

This calls for a way of describing language sounds on an abstract
level. We all hear language-specifically, which means we group \textit{phones}
(speech sounds) into abstract speech units called \textit{phonemes}.
A phoneme is ``the smallest unit of speech that can be used to make
one word different from another word''~\cite{Phoneme_Def}. Different
phones are perceived to be realisations of one phoneme if they fulfill
the same function in a language.

In order to understand the difference between the concept of phones
and phonemes, consider this illustration; All devoted mountaineers
in my homecountry (Switzerland) have come across their fair share
of Edelweiß\footnote{Alpine Edelweiß (\textit{Leontopodium nivale}) is a white flower,
often found in alpine regions. They grow in rock cracks and meadows
at high altitudes.} sightings. Notice that whenever one is sighted (lazier readers will
simply google ``Edelweiß''), we perceive a \textit{realisation}
of Edelweiß. The word ``Edelweiß'' is our class into which we group
sightings of flowers with specific features (white petals, yellow
inflorescences in the middle). This perceptual class is abstract and
stands for a large variety of realisations in nature (some small,
some large, some more fluffy, some smoother). Similarly, phonemes
are abstract units into which we group actual speech sounds (phones).

Boundaries between phonemes are language specific. Take, for example,
the German i\textbf{ch}- and a\textbf{ch}-sound ({[}ç{]} and {[}x{]}
in the phonetic alphabet). These are two physically distinguishable
phones (they are also produced in different ways), but they are generally
perceived as the same language building block (phoneme) of the German
language. Or take the vowels {[}e{]} and {[}i{]}, which are clearly
two different phones (and \textendash{} in germanic languages \textendash{}
two different phonemes). Yet many native Chinese speakers have difficulties
hearing the difference between the two.

In any specific language, phones within one phoneme are meant to be
perceived as representing the same language building block. Infants
must learn the building blocks of their ambient language before they
can meaningfully put blocks (phonemes) together to hear words. Thinking
back to the code breaker in the last paragraph, one might say: the
codebreaker must first learn to decipher the bad handwriting, and
only then can he start making sense of the (now clearly perceived
but still encoded) symbols of the alphabet.

\paragraph{Generalising across speakers}

\begin{figure}
\begin{centering}
\includegraphics[scale=0.18]{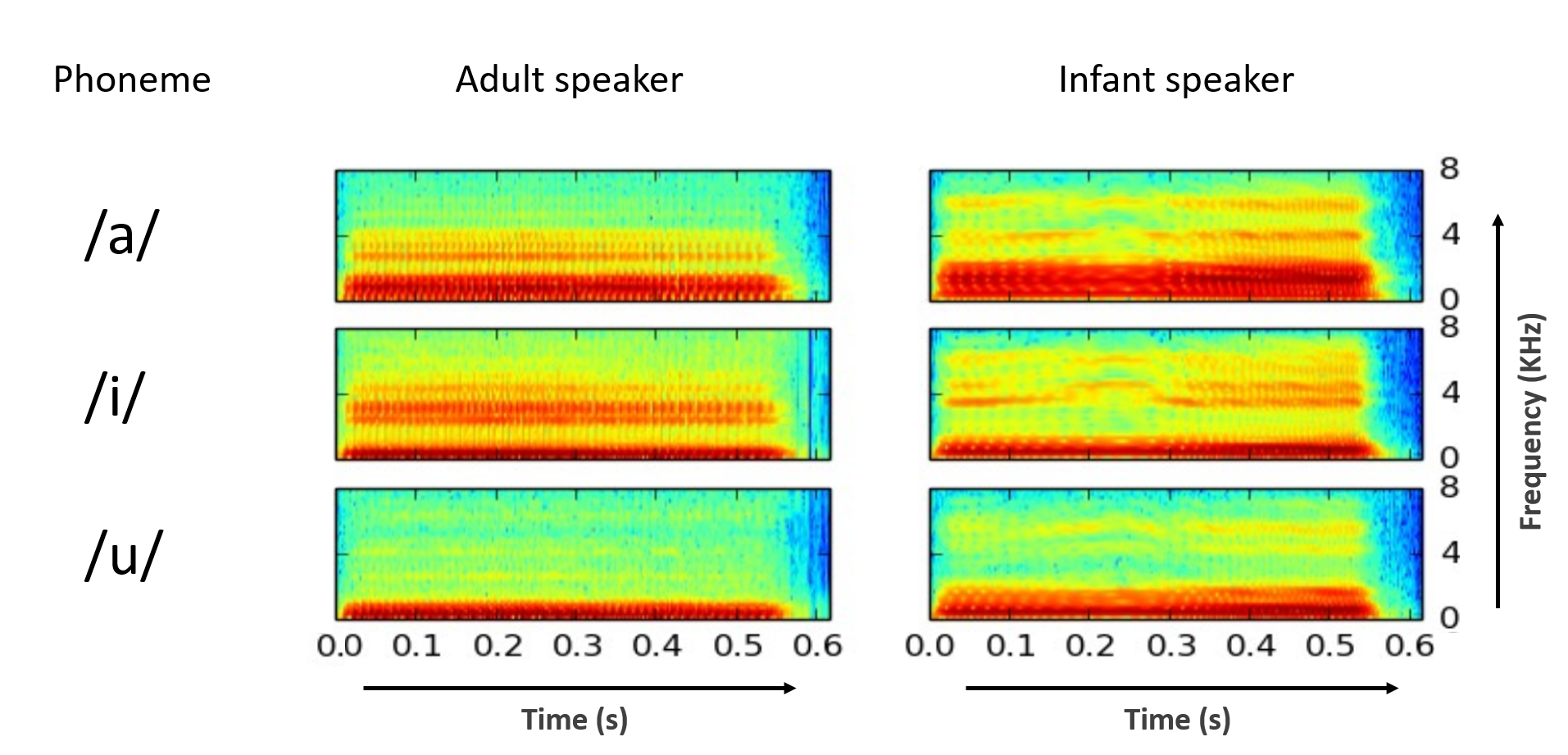}\caption{\label{fig:Spectra_SP_Generalise}Acoustic spectrograms of 3 different
vowels (/a/,/i/,/u/ from top to bottom) from 2 different speakers
(left: adult, right: infant). Colour denotes spectral power (red:
high, blue: low). }
\par\end{centering}
\end{figure}

The variety among phones belonging to one phoneme is especially striking
when we consider speech from different speakers. In Figure~\ref{fig:Spectra_SP_Generalise}
we see acoustic spectrograms of three German vowels /a/, /u/ and /i/.
Note that a spectrogram reflects what we actually physically receive
in terms of sensory information when we hear a sound. While spectrograms
from different vowels do differ (moving up and down between spectrograms
in Figure~\ref{fig:Spectra_SP_Generalise}), so do spectrograms of
the same vowel spoken by different speakers (moving left and right)!
In this case, an adult speaker is compared to an infant speaker when
pronouncing the German vowels /a/, /i/ and /u/\footnote{Vowel sounds were produced using the adult and the infant speaker
in the speech synthesizer \textit{Vocaltractlab}~\cite{P.Birkholz VTL}.}.

This is called the problem of \textit{speaker normalisation}~\cite{Sp_Normalization}
(also sometimes coined \textit{speaker generalisation}). Infants learn
to identify phones that are physically different, but belong to the
same phoneme. ``If my 4-year-old sister says X, it's the same thing
as when my Grandpa says Y.'' \textendash{} X and Y sounding very
different, but, in that specific language, perceived as one and the
same phoneme.

\paragraph{Further difficulties}

Even when uttered by the same speaker, phonemes will have a large
variety of physical realisations. For instance, phones of one phoneme
will vary a lot depending on the \textit{phonetic context}, i.e. where
a phone is placed in a word~\cite{Contextual effects in infant speech perception}.
Also, slow speech results in different acoustic properties than fast
speech~\cite{Slow_Fast_different_phones}. Physical features of phones
are often highly overlapping as we will see later on in section~\ref{subsec:A-series-of}. 

For all vocal gestures in Figure~\ref{fig:Spectra_SP_Generalise}
there was not much pitch alteration (only approx. half a semitone).
Also, the vowels were pronounced alone, without other phonemes before
or after. The reader can well imagine that adding language melody
(varying pitch) and phonetic context to these vowels would make them
even less recognisable.

State of the art computer speech recognition software cannot recognise
phones as belonging to the same class when the speaker, or the rate
of speech or the phonetic context changes~\cite{Computer_speech_recognition}.
And yet, infants are shown to perceive phonetic classes over all these
variabilities~\cite{Contextual effects in infant speech perception,Slow_Fast_different_phones,Kuhl_Dissimilarity}.

\paragraph{Encoding phonetic features}

So how does the human brain manage to generalise so well? Recent work
has been done measuring direct cortical activity from patients undergoing
clinical evaluation for epilepsy surgery. Electrode arrays were implanted
on the superior temporal gyrus (STG), which is attributed to language
perception. The recordings showed phonetic feature encoding, \textendash{}
Neuron clusters in STG responded to spectrotemporal acoustic cues
of the speech sounds~\cite{Mesgarani}. One of these spectrotemporal
acoustic cues is frequently used thoughout this thesis: vowel formants
in formant space (see Box 2). 

However, such findings underline the complexity of the task of language
perception and the fact the infants (who live up to the challenge)
seem to be natural language-learners in a way that is not yet understood.

\begin{figure}
\centering{}\textbf{\large{}}%
\noindent\fbox{\begin{minipage}[t]{1\columnwidth - 2\fboxsep - 2\fboxrule}%
\textbf{Formant space\hfill{}\label{Formant-space}Box~2}

\textit{and phonetic features of vowels}\textbf{\Large{}\medskip{}
}{\Large \par}

{\small{}A concept frequently used thoughout this thesis is formant
space. Below, the German vowel $/a/$ is shown as a waveform (}\textbf{\small{}A}{\small{})
and its corresponding spectrogram (}\textbf{\small{}B}{\small{}).
Prominent (and stable) maxima of the spectrum of a sound are called
formants. Vowels are comparable to musical chords in that each vowel
has a set of formants which together make up its characteristic sound.}{\small \par}

{\small{}In the spectrogram of the German vowel $/a/$, the formant
frequencies are shown (without showing the pitch, which is hard to
distinguish in this plot). Taking the first two formants (}$F_{1}${\small{}
and }$F_{2}${\small{}) we can diametrically plot them in a space
called formant space. This specific realisation of the German vowel
/a/ (author's pronounciation) is drawn in formant space (}\textbf{\small{}C}{\small{},
red arrow) along with formants from a study of American vowels~\cite{Hillenbrand American Vowels}.
(Notice, the American 'short u' vowel is similar to the German vowel
/a/.)}{\small \par}

\smallskip{}

\noindent\begin{minipage}[t]{1\columnwidth}%
\begin{flushleft}
{\small{}\includegraphics[width=1\columnwidth]{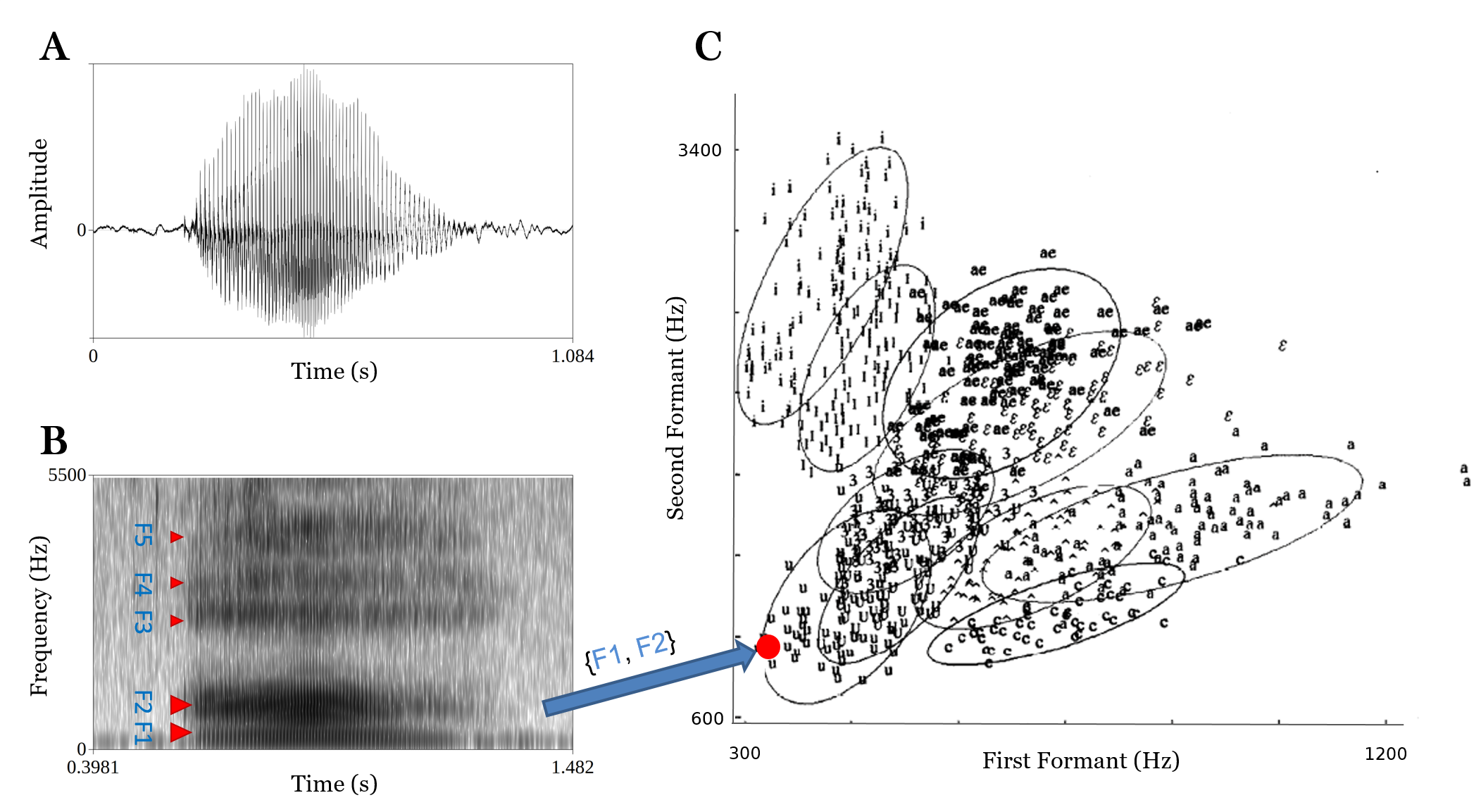}}
\par\end{flushleft}%
\end{minipage}
\begin{flushleft}
{\small{}A and B: Time waveform and spectrogram of the German vowel
/a/ (produced by the author). C: American vowels from many different
speakers plotted in formant space. (Source:~\cite{Hillenbrand American Vowels})}\smallskip{}
\par\end{flushleft}
{\small{}We see that, although formant space is a practical way of
showing (somewhat-) distinguishing vowel properties, vowels are all
but clearly defined. Groups of different vowel formants strongly overlap.
Keeping vowels apart by looking at their physical features (e.g. formants)
is not simple.}{\small \par}

{\small{}Additionally, measuring formants is not at all trivial~\cite{Formant Evaluation Comparison}.
In this thesis all formants were evaluated using the Burg method~\cite{Burg Method}.}{\small \par}

{\small{}Whenever one formant value is shown for a certain phone from
a certain speaker, the medium was taken over all formant values which
were extracted from the vowel over time.}\smallskip{}
\end{minipage}}
\end{figure}

\medskip{}
We conclude that, in order to learn speech, an infant must first learn
to cluster phones into groups of phonemes, which in turn fulfill a
specific function in that language. A newly-born is thought to listen
to ambient speech and train its own auditory system to do this ``categorisation''
correctly. Only then, eventually, can a sequence of speech sounds
be loaded with actual meaning. And, moving on, knowing how phones
\textit{should} sound is crucial for starting to imitate them.

\subsubsection{The challenge of speech imitation}

We have seen some challenges that infants face when learning to perceive
language throughout the previous section. In general, getting your
perception right is crucial for imitating correctly. This is the underlying
assumption I make throughout my thesis. We will, however, discuss
this point again in Section~\ref{CaregiverModel}.

The afore mentioned difficulties involved in the perceptual aspect
come along with the staggering challenge of actually imitating speech
sound. We understand how difficult playing the violin is, not by hearing
a violin solo, but by investigating sound production on a violin itself.
In this section I'd like to introduce the 'instrument' by which we
produce speech and then go on to a phase in child development called
'babbling', in which infants are thought to learn to control this
speech instrument \textendash{} their own vocal tract.

\paragraph{The Vocal Tract}

\begin{figure}[H]
\centering{}\includegraphics[width=1\columnwidth]{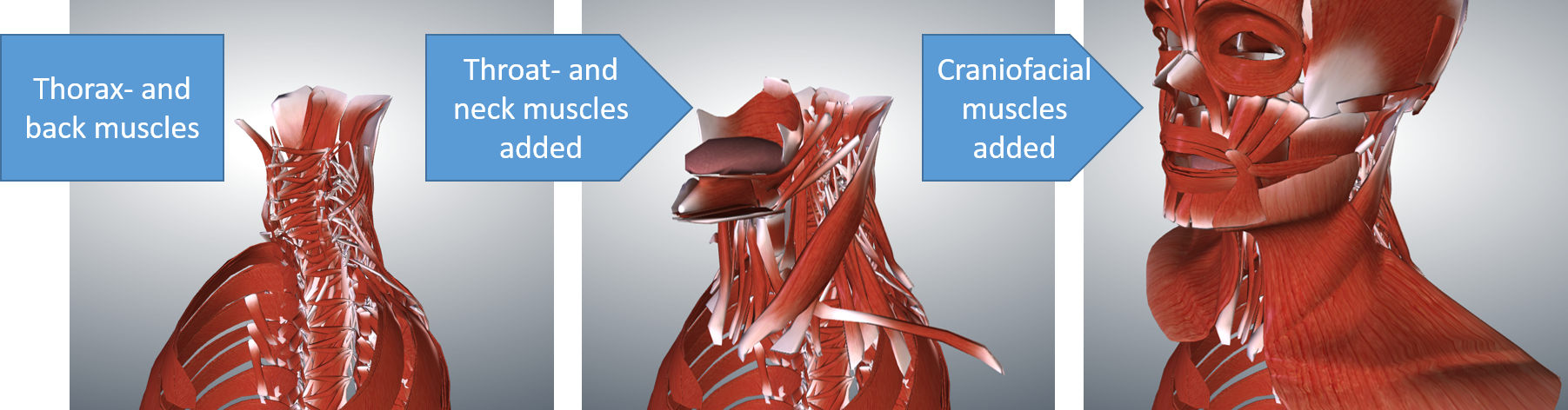}\caption{\label{fig:Vocal Tract}The muscles of the human vocal tract, shown
in different layers. Image created using software freely available~\cite{Articulators}. }
\end{figure}

The vocal tract is comprised of all parts actively involved in speech
production. Figure~\ref{fig:Vocal Tract} illustrates the complexity
of the human vocal tract. In order to produce speech, approx. 100
muscles are simultaneously involved~\cite{100DimUnique}. Using this
highly complex subsystem to articulate a phrase would correspond to
a 100-dimensional trajectory, where a certain motor configuration
would have to develop with precise timing (thus a trajectory). The
figure shows various layers of muscles: 
\begin{itemize}
\item thorax and back muscles, which are important for lung contraction
and airflow production,
\item throat and neck muscles. These include the muscles around and between
the cartilage that makes up the larynx. In the larynx ('voicebox')
the vocal folds are stretched/relaxed, and the glottis (the opening
between the vocal folds) is made wider/narrower. When air passes the
(nearly closed-) vocal folds they begin to vibrate, thus creating
a tone which is then 'shaped' in the rest of the vocal tract,
\item craniofacial muscles, together with the tongue, are mainly responsible
for articulation of speech sounds.
\end{itemize}
In principle, every one of these muscles has to be controlled by the
human brain. However, we can simplify the problem by neglecting most
of these muscles and focusing on only a few. In phonetics, it is common
to look at a handful of\textit{ articulators}, which include the lips,
teeth, alveolar ridge, hard- and soft palate, pharynx and larynx (see
Figure~\ref{fig:Vocal Tract-1}, left). The tongue itself is often
labeled according to its segments: The tip, blade, front, back or
root of the tongue.

\paragraph{Infant babbling}

Having given an impression of our 'voice organ', the instrument that
has to be explored and learned by the infant learner, we now move
on to consider \textit{how} the infant learns to articulate speech
with his own vocal tract.

Infants seem to explore their own capability to produce vowels (cooing,
around 3 months) and strings of simple syllables (around 7 months)~\cite{Book_Language_Development}.
This process, which goes on until they are able to produce recognisable
words (normally at around 12 months) is called \textit{babbling}.
Strings of syllables produced around 7 months often include small
alterations over time (for example ``baba'', ``bebe''). This specific
phase is generally known as\textit{ canonical babbling. }Babbling
seems to be present in all children acquiring language and has been
studied in infants across the world. An intriguing aspect of babbling
is that it does not need to be connected to the infant's emotional
state. Infants seem to babble spontaneously and incessantly when they
are both emotionally calm or upset/excited~\cite{Book_Psychology_of_Language}.

The universal nature of infant babbling might suggest that infants
use exploration (of their own vocal tract's ''states'') and, reinforcing
those lucky vocal gestures which actually sounded 'right', actually
begin to imitate their caregivers. Research shows that, when babbling,
infants do match patterns of rhythm of the language(s) to which they
are exposed~\cite{Babbling_Language_specific}. They use intonation
patterns (as in Box 3) and timing of surrounding speech, mainly using
the consonants and vowels that occur most frequently in the caregiver's
language~\cite{Babbling_Language_specific}.
\begin{figure}
\textbf{\large{}}%
\noindent\fbox{\begin{minipage}[t]{1\columnwidth - 2\fboxsep - 2\fboxrule}%
\textbf{Are young infants imitation-driven?\hfill{}}{\small{}\label{Box-3:Imitation}}\textbf{Box
3}\textbf{\small{}\medskip{}
}{\small \par}

{\small{}The subject of imitation is debated among child researchers.
It is not yet clear whether actions of infants in their first months
are truly motivated by the desire to imitate or not. Studies on infants
younger than 3 weeks showed that these sometimes do match adult behaviour,
but to some it seems unlikely that this matching is imitation-driven~\cite{Imitation}. }{\small \par}

{\small{}Studies on tongue protrusion (among other facial gestures)
showed above-baseline activity, when the caregiver also stuck out
his/her tongue (e.g.~\cite{Meltzoff Moore Imitation}). But is increased
tongue protrusion the result of imitation, or simply of the infant's
arousal when it sees something interesting (the caregiver sticking
out his/her tongue)? }{\small \par}

{\small{}Recent meta-analysis shows poor statistical power for some
studies that speak }\textit{\small{}against}{\small{} facial imitation
and draws attention to the fact that in some cases the experimental
setup choked the motivation to imitate~\cite{Metastudy_Imitation}.}{\small \par}

{\small{}We know that infants' attempts at producing speech strongly
resemble ambient speech. Figure~\ref{fig:Cry} shows typical French
vs typical German infant crying. The ambient French speech intonation
patterns being predominantly upward, the infants (who hear ambient
speech from the 6th month of pregnancy on) seem to have the same upward
intonation pattern when crying (and vice versa with German infants)~\cite{Cries_French_German}.}{\small \par}

{\small{}Throughout this thesis, I assume infants are driven by the
desire to imitate, also when learning speech. In the discussion section,
I will compare this underlying assumption and look at models of speech
acquisitions that take different stances.}\smallskip{}

\noindent\begin{minipage}[t]{1\columnwidth}%
\begin{center}
{\small{}\caption{{\small{}\label{fig:Cry}Time waveform and spectrograms of a typical
French cry and a typical German cry. Source: \cite{Cries_French_German}}}
\includegraphics[width=1\columnwidth]{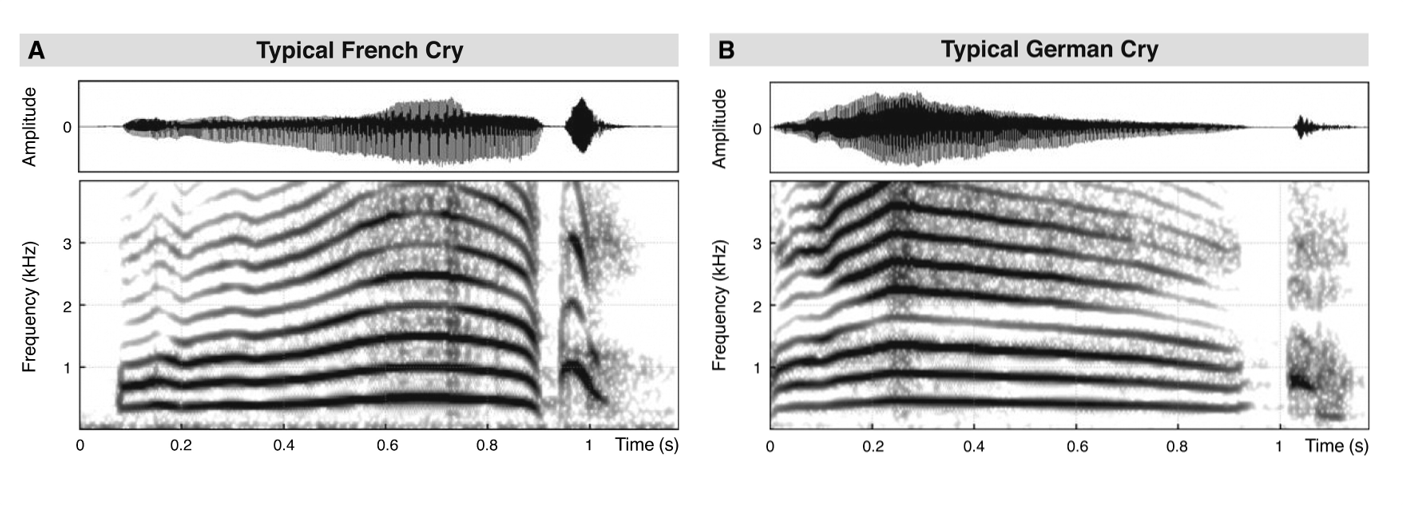}}
\par\end{center}%
\end{minipage}%
\end{minipage}}{\large \par}

\bigskip{}
\end{figure}

\begin{figure}
\includegraphics[width=1\columnwidth]{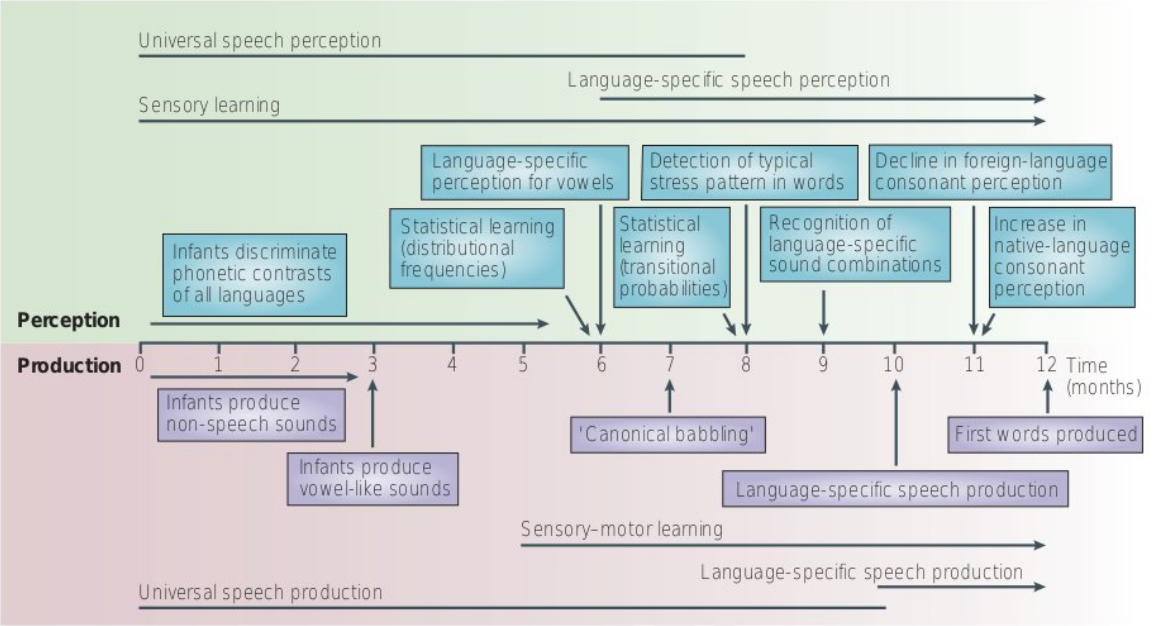}\caption{\label{fig:The-development-of}The development of speech perception
and speech production in typically developing infants during their
first year of life. Taken from \cite{P.Kuhl Early language acquisition}.}
\end{figure}

It is important to see that babbling is not the only stage important
for language learning. In Figure~\ref{fig:The-development-of} canonical
babbling is represented as one stage in the context of a whole range
of production- (and perceptual) stages. Some of these activities seem
to happen simultaneously. Humans seem to be equipped with a diverse
set of abilities which, combined and rightly timed, on the perceptual
side as well as on the production side, contribute to the acquisition
of language and speech. 

\subsubsection{\label{subsec:Infant-directed-speech}Infant-directed speech}

\paragraph{Motherese}

\begin{figure}
\centering{}\includegraphics[width=0.8\columnwidth]{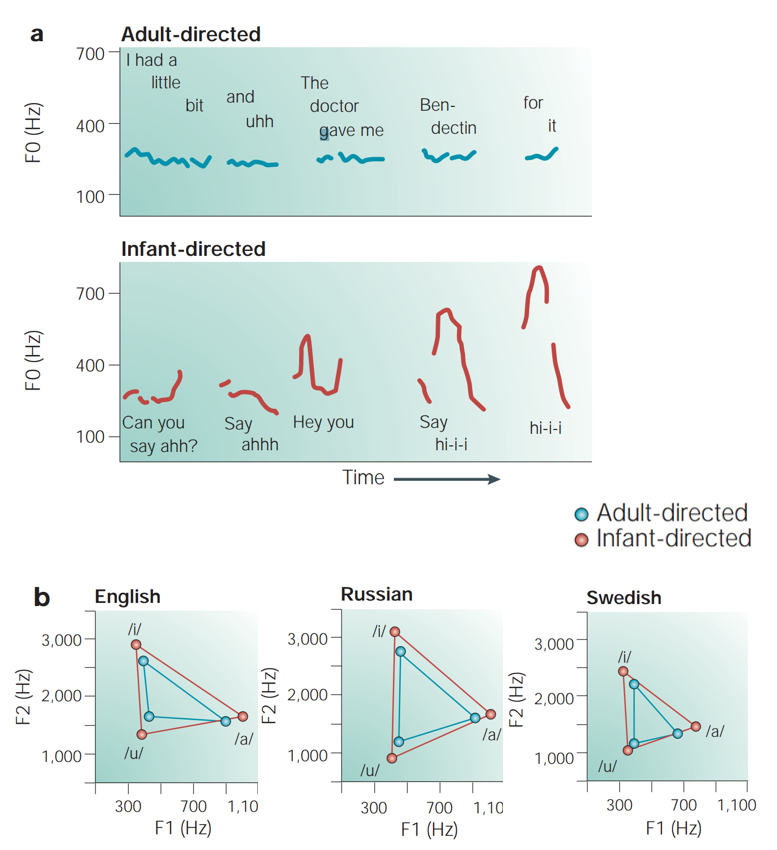}\caption{\label{fig:motherese_intro}\textbf{a}: Characteristic pitch contours
of adult-directed speech and motherese (\textit{see caption}). Large
pitch variation is typical for infant-directed speech.\protect \linebreak{}
\textbf{b}: Vowel triangles for adult-directed and motherese speech
in formant space from studies in three different languages. All three
studies show increased vowel-space for motherese speech (Vowels are,
formant wise, more distinguishable).\protect \linebreak{}
Taken from~\cite{P.Kuhl Early language acquisition}.}
\end{figure}

Another source of help for the young learner is deliberate infant-caregiver
interaction. When we talk to infants and young children, we use a
special 'mode' of speech called \textit{motherese}. Caregivers in
most countries use motherese when addressing children. The fact that
motherese is a common phenomenon suggests that infants need infant-directed
speech to learn language. When addressing infants, we tend to express
ourselves with a lot of change in pitch (see Figure~\ref{fig:motherese_intro}~a).
Also, motherese is typically slower and better articulated. An important
aspect in vowel pronounciation is the larger vowel space spanned by
the formants when infant directed (see Figure~\ref{fig:motherese_intro}~b).
Studies among English, Russian and Swedish mothers show that the medium
formants of their vowels are further apart (e.g. /u/ has slightly
lower $F_{1}$ and $F_{2}$ values while /i/ has lower $F_{1}$ but
higher $F_{2}$ values for motherese speech, compared to adult directed
speech).

In Figure~\ref{fig:motherese_intro}~b triangles are drawn by connecting
positions of $/u/$, $/i/$ and $/a/$ in formant space. Those three
vowels act as extreme vowel sounds. Other vowels are articulated somewhere
inside the triangle spanned by $/u/$, $/i/$ and $/a/$. This is
why phoneticists use the concept of vowel triangles. Larger triangles
mean that all vowels (the corner points $/u/$, $/i/$ and $/a/$,
but also those inside the triangle \textendash{} $/e/$ and $/o/$
for example) will be further apart and more easily distinguishable.

\paragraph{Caregivers imitate}

Infant directed speech (i.e. motherese) is thought to aid the young
learner in the perceptual task. But recent research suggests that
parent-child interaction might also directly contribute to the learning
process (not merely on the perceptual side)\footnote{See ``Creating the cognitive form of phonological units: The speech
sound correspondence problem in infancy could be solved by mirrored
vocal interactions rather than by imitation'', section 4 (Support
for Mirrored Equivalence). Full citation: \cite{Messum}}. We naturally think of imitation being done by the infant (infants
imitate their parents). But mothers frequently imitate their infants'
speech \textendash{} in fact more often than the other way round (e.g.~\cite{Palby_Imitation},
and other studies quoted in~\cite{Messum}). This imitation might
form associations between the learner's speech gesture and the caregiver's
imitatory response. The infant learner might, for example, execute
a certain motor pattern which (by chance) produces something similar
to an actual speech sound. In response, the caregiver imitates and
produces a correct pronounciation of that specific speech sound. In
the process, the infant might form associations between the 'motor
pattern' which was executed in order to perform the gesture and the
auditory signal received from the caregiver (the correct pronounciation).
This motor pattern would then be reinforced and used again (maybe
with a slight variation / exploration).

For now let it be said that the role of the caregiver, specifically
when using infant directed speech, seems crucial to language acquisition
of infants. But \textit{Listen and Babble} works without direct caregiver-infant
interaction while the model learns to articulate. I argue that this
could be the reason for certain problems\textit{ Listen and Babble}
faces.

\newpage{}

\subsection{\label{subsec:Model_Explained}Listen and Babble \textendash{} A
model of vowel acquisition through imitation}

My research builds upon Murakami's work, described in his thesis~\cite{Thesis Murakami}.
In short, Murakami successfully simulated vowel acquisition using
machine learning. For a detailed description of his work I again refer
to~\cite{Thesis Murakami}. I shall describe the basics in this section,
but also try to show some points where the model still needs improving.

\subsubsection{Basic concepts}

The model of vowel acquisition through imitation: \textit{Listen and
Babble} seeks to understand how speech is learned by infants. Our
brains work with rewarding signals (e.g. dopamine\footnote{To be precise: Dopamine codes for the so-called 'prediction error',
the discrepancy between prediction and actual reward.}) and ``use them for influencing brain activity that controls our
actions, decisions and choices''~\cite{Scholar_Reward_signals}.
Predictions are then made, based on rewards in the past, and prediction
errors computed. These prediction errors then help to form new (and,
over time, better) predictions for the future. These predictions are
strongly linked to actions. In speech, for example, the prediction
would be the actual sound produced by executing a combined position
of the articulators. Like this, the infant would learn to predict
the outcome (sound) of its own actions (vocal gesture) with steadily
increasing precision. Muscle movements (or motor patterns in the cortex)
are mapped to the infant's own speech sounds: The young human learns
to speak. 

Before introducing the full model \textit{Listen and Babble}, we must
first cover some key concepts of the model.

\paragraph{Machine learning}

Learning new actions based on previous outcomes is central in a field
called \textit{machine learning. }Machine learning seeks to produce
increasingly 'correct' actions from a machine (e.g. a programmable
robot) without constantly giving explicit instructions to the machine.
There are various ways of doing this, each suitable for a certain
range of applications. For now, let us look at two areas of machine
learning\footnote{For a more detailed (and valuable!) description of machine learning,
explained in context of the \textit{Listen and Babble} model, see
\cite{Thesis Murakami}, pages 12-20.}:\\
\linebreak{}

\begin{tabular*}{1\columnwidth}{@{\extracolsep{\fill}}>{\raggedright}p{0.45\columnwidth}>{\raggedright}p{0.45\columnwidth}}
\textbf{\large{}\medskip{}
}\textbf{\small{}Supervised Learning}{\small \par}

{\small{}Let's assume we are given a set of data. Each set can either
belong to a category or produce some state. The task is to programme
a computer to learn to make accurate predictions of the label (category)
or state if a new data set is presented to it. In order to make good
predictions, the programme has to be trained using sets of data that
already are labeled, or where the produced state is known. The learner
is 'supervised' in that each training set is pre-labeled. Only after
training, when unlabeled datasets are presented, does the trained
programme make its own prediction.}\textbf{\large{}\medskip{}
} & \textbf{\large{}\medskip{}
}\textbf{\small{}Reinforcement Learning}{\small \par}

{\small{}Reinforcement Learning uses a weaker form of supervision:
the learner (agent) is given neither goal nor directions by a human
supervisor. The RL-agent must be programmed to interact with its }\textit{\small{}environment}{\small{},
which in turn must:}{\small \par}
\begin{itemize}
\item {\small{}inform the RL-agent about its current state,}{\small \par}
\item {\small{}yield a scalar reward for the last action.}{\small \par}
\end{itemize}
{\small{}This is repeated many times {[}RL-agent performs action,
receives state and reward information from the environment{]}. During
many iterations a RL-agent typically tries to maximise future rewards
by adjusting its action policy based on the (cumulative) information
from the environment.}\textbf{\large{}\medskip{}
}\tabularnewline
\end{tabular*}\textbf{\large{}\bigskip{}
}{\large \par}

Both areas are relevant for \textit{Listen and Babble;}
\begin{enumerate}
\item Supervised learning is used to train an auditory system to correctly
classify speech sounds, while 
\item Reinforcement Learning is used by a computer simulated infant learner
to acquire the ability to reproduce (or imitate) such speech sounds.
\end{enumerate}
In the next paragraph, I attempt to explain the full model since my
research also uses the same model.

In some of the following sections, while summarising how \textit{Listen
and Babble} works, its achievements and prevailing difficulties, I
will, on occasion, be using words and concepts that are only introduced
in section~\ref{sec:Methods} (Methods). In such cases I ask my reader
to hang on and 'live with the gaps', or skip to the relevant sections
before returning again to these.

\subsubsection{The model}

I also ask the reader to recall that we looked at two challenges (or
stages) in infant speech aquisition, namely: perception and imitation.
The \textit{Listen and Babble} model works on the assumption that
speech perception (right hearing of speech sounds) is aquired \textit{before}
the infant actually imitates. This would mean that we can treat these
two separately. Murakami, in his model \textit{Listen and Babble},
first trained an \textit{Auditory System} to accurately categorise
speech sounds as one of 4 classes:
\[
/a/,/i/,/u/,\,null
\]

The vowel classes signify German vowels and 'null' class simply stands
for neither $/a/,/i/\,or\,/u/$. Computer-synthesised speech samples
were produced with VocalTractLab (explained in section~\ref{subsec:VocalTractLab}),
then individually listened to and labeled by hand with either $/a/,/i/,/u/$
or $null$. The auditory system was trained using \textit{Supervised
Learning}. Remember that in Supervised Learning, the learner is trained
using many correctly labeled sets of data and learns to make its own
predictions for new, unlabeled sets of data{\small{}}\footnote{{\small{}Take, for example, the task of predicting the share value
of a certain firm. Here, the state is simply the value of the share.
If we mined numbers of certain twitter keywords mentioned in connection
with that firm, we might use this 'twitter-mood', together with the
current share value, to predict the next value of the share. In this
case, twitter mood + current share value for a certain timepoint is
one dataset. The following share value (say, 1 day afterward) is the
state. Take many such datasets + states (e.g. from the last 3 years)
and train a computer to make accurate predictions for future states.}{\small \par}

{\small{}These predictions should be reasonably reliable, and the
said firm has a good early-warning system (assuming a stable market!).}{\small \par}}. In this case, the dataset consists of sound files (labeled by the
human user), and the auditory system is trained to accurately classify
speech sounds.

Training the Auditory System models the perceptual stage, where infants
learn from ambient speech to recognise language-specific speech sounds
and learn to form the same linguistic boundaries (e.g. between vowels)
as their caregivers would have. The second stage (imitation) is done
using \textit{Reinforcement Learning}.

Recall that infants seem to explore their own speech-instrument (vocal
tract), trying out all kinds of gestures and reinforcing such which
yield 'promising' speech sounds \textendash{} sounds which sound 'right'
in the infant's linguistic setting (parents, siblings, local communities,
etc). The Reinforcement Learning agent (RL-agent) forms a vocal gesture,
uses a synthesiser to form an output (speech sound) which then in
turn is analysed by the (already trained) auditory system. The auditory
system then returns a reward signal to the RL-agent: high rewards
if the speech sound was recognised as either $/a/,/e/,\,or\,/u/$,
low rewards if the sound wasn't recognised (class $null$).

Imagine a human infant babbling (forming its own vocal gesture), then
hearing its own speech sound. If the sound was recognised as similar
to the caregiver's speech, the infant will be delighted and continue
to form similar vocal gestures.

The RL-agent interacts with (passes on articulator positions to) a
simulated vocal tract. The vocal tract creates speech sounds using
those articulator positions. The speech sounds are then perceived
by the auditory system (which has already been trained to perceive
correctly!). The auditory system passes a confidence vote, something
like: ``How much did this sound really sound like an $/a/$?'' A
reward is computed from the confidence vote and the RL-agent updates
its policy based on the reward and executes a new vocal gesture by
sending new parameters to the vocal tract. This is repeated ad infinitum,
or until rewards become satisfactory (we have learned the given target
vowel).

\begin{figure}[H]
\includegraphics[width=1\columnwidth]{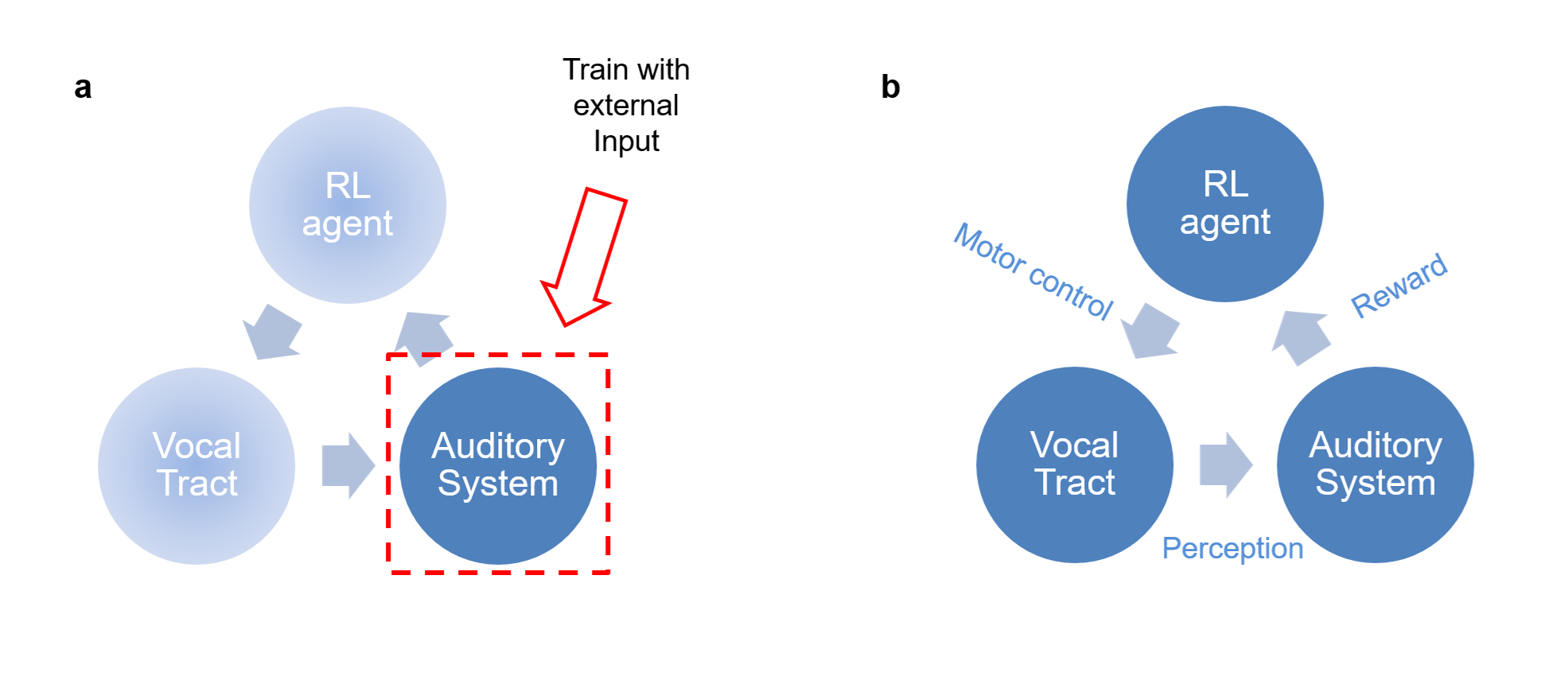}\caption{\label{fig:ListenandBabble_Model}\textbf{a}: Perceptual stage of
\textit{Listen and Babble}. The auditory system is trained with external
input (hand-labeled speech samples).\protect \\
\textbf{b}: The imitation stage of \textit{Listen and Babble}. The
RL-agent passes a set of motor parameters to the Vocal Tract (the
VTL synthesiser) which in turn produces the resulting speech sound.
This sound is listened to by the (already trained!) auditory system
and a reward is passed back to the learner for each set of motor parameters.
This process continues until a reward boundary is achieved for a targeted
vowel.}
\end{figure}

The model's two stages (perception and imitation) are illustrated
in Figure~\ref{fig:ListenandBabble_Model}. The synthesiser (vocal
tract model), the auditory system and the RL agent / algorithm are
discussed in more detail in sections~\ref{subsec:VocalTractLab},
\ref{subsec:Auditory-System} and \ref{subsec:RL-algorithm}.

\subsubsection{\label{subsec:Achievements}Some results}

\begin{figure}[H]
\includegraphics[width=1\columnwidth]{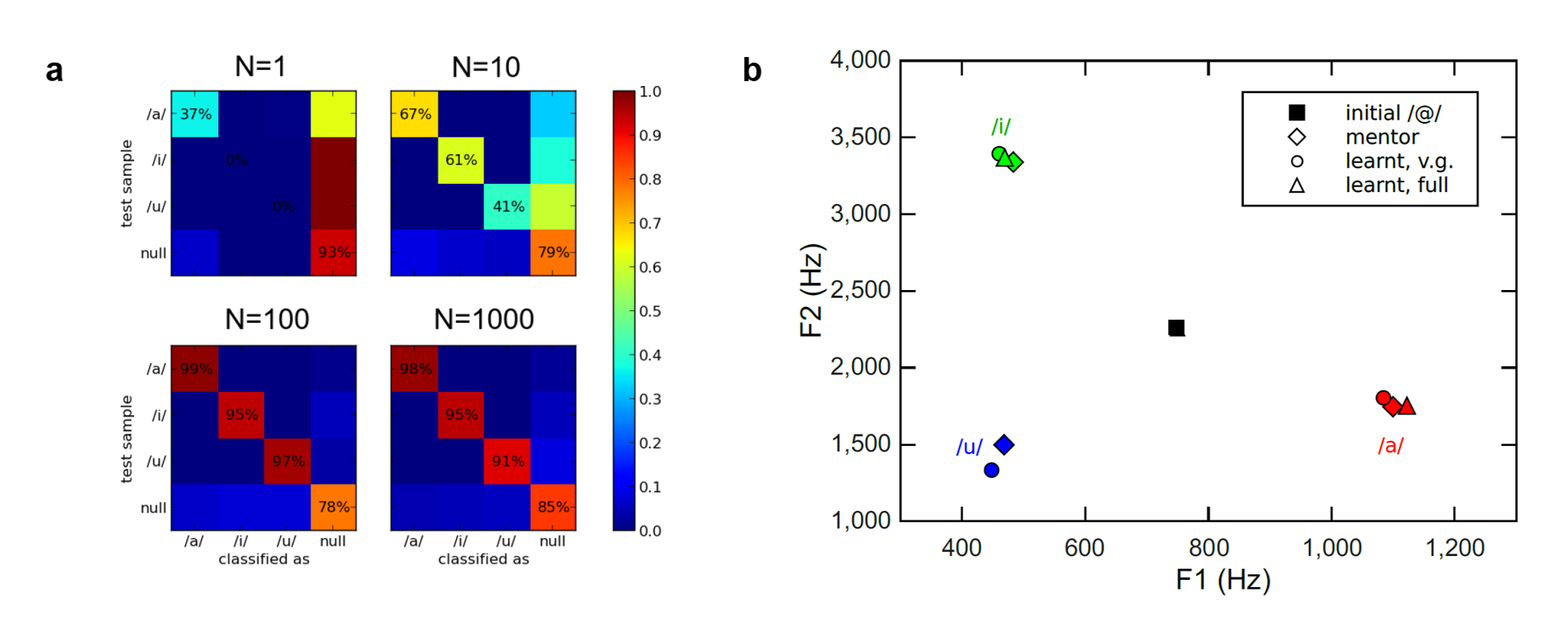}\caption{\label{fig:Imitation_Model}\textbf{a}: Confusion matrix of an ESN
of $N_{Neurons}=\{1,10,100,1000\}$. A confusion matrix shows what
fraction of testing samples of each class are classified as that class.
Few vowels were classified as null class (or other vowels), however
15\% of null-speech sounds were (wrongly) classified as one of the
vowels. An ESN manages to correctly classify all 4 classes $/a/,/i/,/u/,\,null$
with an accuracy of approx. 94\% when using more than 100 neurons.\protect \\
\textbf{b}: Imitation learning in formant space. The agent starts
with 'neutral' articulator positions yielding $/@/$ sounds (\textit{square})
and then moves on to imitate the mentor vowels (\textit{diamond}),
producing acoustically similar vowel versions when learning with all
parameters (\textit{triangles, \textendash{} }only $/i/$and $/a/$learned)
or visually guided 13 parameter learning (\textit{circles}).}
\end{figure}

Murakami managed to sucessfully train the Auditory System to classify
$/a/,/i/,$ $/u/$ and $null$ with an accuracy of approx. 94\%. Part
of the Auditory System involved in training is the \textit{Echo State
Network} (an Artificial Neural Network). For ESN reservoir sizes (number
of artificial neurons) of over 100, the error rate plateaus. (An untrained
ESN would, on average, have a random classification accuracy of around
24\%). In Figure~\ref{fig:Imitation_Model}~a, the confusion-matrix
is displayed for an Echo State Network of 1000 neurons.\\

In a second step a RL-agent learned to imitate vowels $/a/,/i/$ and
$/u/$. Remember the complexity of finding the right position for
each articulator (tongue, jaw, etc.) in order to produce a correct
phoneme? In \textit{Listen and Babble}, a 16-dimensional motor space
is explored\footnote{This is not uncommon for Reinforcement Learning problems, which frequently
deal with huge numbers of possible actions (or \textit{large action-spaces}). 

On occasion, a reinforcement learning problem will enable a RL-agent
to visit all possible states multiple times, yield a (sometimes delayed)
reward for each of these states, and compute a 'value' of that state,
or state-action pair (``how much reward we might in future expect,
when we're in that state and perform this action''). But often, the
RL-agent will only be able to visit a few of these states, simply
because there are so many. Take, for example a RL-agent learning to
play chess. There are approx. $10^{55}$possible ways to place the
chess pieces on the board! In that case, a Reinforcement Learning
algorithm would have to cope with the fact that they will only be
able to explore a very small number of possible states. This is also
so in the state space of a human vocal tract.} in order to find vowel-producing positions of the articulators (more
about the articulator parameters in section~\ref{subsec:VocalTractLab}).

In the case of $/a/$ and $/i/$, all 16 motor parameters (articulator
positions) were learned, each articulator starting from an initial,
neutral position (the position of the articulators when producing
German phoneme $/@/$ as in \textit{'viel}\textbf{\textit{e}}'). One
important finding was that the model couldn't learn $/u/$ with all
16 degrees of freedom. The learner needed to keep jaw and lip parameters
(3 in total) preset with the 'mentor parameters' (the known parameters
from a standard VocalTractLab speaker). The reason is that $/u/$
requires quite extreme lip and jaw positions, which are simply not
found by the RL-agent. The fact, however, that the learner needs to
know lip and jaw positions (and not derive them from exploration only),
is not too far-fetched. The lips' shape is a conspicious visual feature
in the face of a caregiver, which could easily be picked up by the
infant and associated with the simultaneous speech sounds. Thus, visually
guided learning (13 motor parameters) proved sucessful when learning
all three vowels, including $/u/.$

Figure~\ref{fig:Imitation_Model} provides us with a visualisation
of the correct-classification rates of the auditory system (\textbf{a})
and the quality of the learned vowels compared to mentor vowels in
formant space (\textbf{b}) for learning with either all parameters
(16) or visually guided learning (13 parameters).

\subsubsection{Open questions}

In this section, I will try to show a selection of questions that
were still left open in the original \textit{Listen and Babble} model.
I specifically mention these in order to show how my work takes up
these questions, rules out certain approaches and offers further suggestions.

\paragraph{Speaker Normalisation}

Just like other models dealing with speech perception, \textit{Listen
and Babble} faces the problem of \textit{speaker normalisation}. When
the infant's auditory system was only trained on adult speech samples,
its own speech was not recognised (we discussed this challenge of
speech perception in section~\ref{subsec:The-challenge-of}). So
that the model would work, the auditory system was trained on both
infant and adult speech. But obviously an infant would not have the
opportunity to perceptually train on correctly pronounced vowels from
his own vocal tract! 

We could however argue that the \textit{Listen and Babble} model should
be applied to speech acquisition in a social context. We might interpret
the auditory system not as that of the infant itself, but of the (fully
trained, and already able to generalise across speakers) caregiver
auditory system. The rewards would then no longer be interpreted as
only inner nerve and hormone signals but as social feedback (approving
sounds, \textendash{} rewards given by the caregiver).

In my opinion, however, this interpretation forfeits some of the advantages
of this model over other models of speech acquisition (most importantly
the fact that \textit{Listen and Babble} offers an explanation as
to why infants also take to babbling without social stimulus, or direct
caregiver influence~\cite{Book_Psychology_of_Language}). It makes
sense to think of the auditory system as being the infant's, giving
rewards for speech-like sounds without needing the ever-present interaction
with a caregiver.

In this thesis, I sought to make some first steps in two directions:
\begin{itemize}
\item How well does speaker generalisation work in connection with echo
state networks (part of the auditory system)?
\item How will learning be affected by an auditory system trained to generalise
over a range of ages and both genders?
\item Should we change our approach (two-stages, target acquisition before
imitation) in order to tackle the problem of learning speech generalisation?
\end{itemize}

\paragraph{Small steps toward language}

\textit{Listen and Babble} only learns 3 very distinct vowels. Those
vowels, $/a/,/i/,/u/$ are the most distinct German vowels, so it
seemed natural to start with these and it obviously is just the first
of many steps toward actually perceiving and imitating language. One
might ask: What if vowels that are phonetically more similar were
introduced? Will the auditory system still be able to distinguish
well? What effect would eventual lower sensory accuracy have on learning?

In this thesis, I introduced two further vowels ($/o/$and $/e/$,
which are similar to $/u/$ and $/i/$respectively) and sought to
make the project implementation more flexible for even implementing
arbitrary speech gestures\footnote{See Readme.md in the \textit{Listen and Babble} repository, ``Towards
arbitrary vocal gestures''~\cite{Repository}.}.

\cleardoublepage{}

\section{\label{sec:Methods}Methods}

\textcompwordmark{}

In this section, I describe the tools used in the model \textit{Listen
and Babble}, focusing on changes I made in my research. For a more
detailed description of the reinforcement learner, I refer to Murakami's
thesis.

\subsection{\label{subsec:VocalTractLab}VocalTractLab}

\begin{figure}[h]
\includegraphics[width=1\columnwidth]{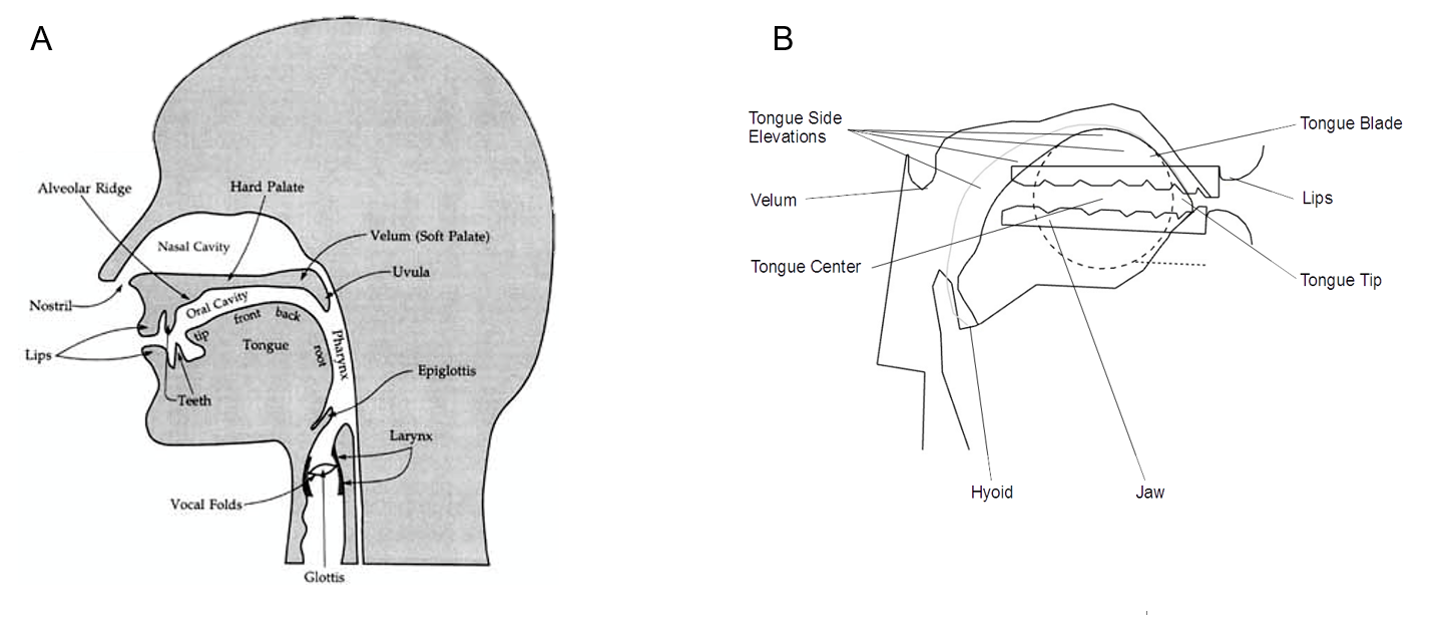}

\caption{\label{fig:Vocal Tract-1}\textbf{A}: Articulators of the human vocal
tract. Taken from~\cite{Labeled vocal Tract}.\protect \linebreak{}
\textbf{B}: 2-dimensional cut through VTL's vocal tract model.}
\end{figure}

\textit{Vocaltractlab} (VTL) is a state-of-the-art speech synthesiser~\cite{P.Birkholz VTL}.
VTL produces speech sounds based on a 3-dimensional vocal tract model.
The model can be controlled using 20 articulator coordinates. Once
a certain articulator configuration is set by the user, the airflow
through the open space (which is given by the position of the articulators)
is simulated. This airflow consists of pressure waves produced by
a model of vocal fold motion. Solving the corresponding differential
equations of air passing the hyoid, velum, being guided by the tongue
and lips results in a life-like (though artificial) speech sound.
A 2-dimensional cut through VTL's vocal tract model is shown in Figure~\ref{fig:Vocal Tract-1}\textbf{b}.
The software comes with a GUI (Graphical User Interface), where the
user can hand-position the articulators and directly produce the corresponding
speech gesture. 

Most speech synthesisers produce sound based on speech recordings.
VTL is different in that the (somewhat simplified) process of speech
production in the human vocal tract is simulated. This is convenient
for our cause, since we can enter a set of articulator parameters
(or \textit{motor parameters}) and VTL produces a speech sound based
on those parameters.

The Reinforcement Learner uses VTL as a part of its environment. 2
of the 20 parameters controlling the articulators are constant for
German vowels (velic opening and horizontal jaw position) and thus
will not be learned. Two more parameters (tongue root coordinates)
also don't count as learning parameters since they are derived from
other motor parameters. The remaining 16 parameters can be individually
explored by the RL-agent by simply giving a set of parameters to VTL,
which synthesises a speech sound that may or may not sound like a
German vowel. For such applications, VTL comes with a Linux-API callable
in Python and Matlab where VTL can be called directly in any given
Python/Matlab code.

The remaining 16 degrees of freedom constituting the motor space are:
\begin{itemize}
\item 10 tongue parameters: tongue body coordinates (TBX, TBY), tongue tip
coordinates (TTX,TTY), tongue center coordinates (TCX, TCY), and tongue
side elevations (TS1, TS2, TS3, TS4),
\item 2 lip parameters: lip separation distance (LD) and lip protrusion
(LP)
\item 2 hyoid coordinates (HX, HY)
\item jaw opening angle (JA),
\item velum shape (VS).
\end{itemize}
(List taken from~\cite{Thesis Murakami}).\bigskip{}

VTL comes with two preconfigured speakers (one male infant and one
male adult speaker). VTL speakers are anatomically constructed using
MRI data from human vocal tract sections. One version of VTL (which
is still in development) allows the user to interpolate speakers of
various ages between infant and adult stage. I made use of this feature
when producing ambient speech from multiple speakers of various ages
(see section~\ref{subsec:A-series-of}).

These 16 parameters are limited by the anatomy of the current VTL-speaker.
Each parameter will have a different physiological range of possible
values (e.g. the jaw will not be able to open more than a certain
value which is still considered realistic). We can, however, normalise
the parameters to {[}0,1{]}. Motor learning will thus take place in
a 16-dimensional cube with edge length 1 (range {[}0,1{]} for each
edge).

For both (infant + adult) speakers, vocal tract configurations are
available for $/@/,/a/,/i/$ and $/u/$ in VTL 2.1. These were fitted
to actual MRI data of patients while they were speaking.

We call these sets of parameters mentor or target parameters. I hand-set
these target parameters for all non-standard speakers from age 0 to
20. These hand-set vowels (which I call \textit{prototype vowels})
are not based on MRI data, but rather interpolated from vowel gestures
of the standard VTL speakers.

A point worth mentioning here, especially in the context of trying
to hand-fit speaker anatomies and gestures in VTL, is this: mapping
parameter configuration to speech sound is not unambiguous in both
directions. Many different articulator positions can produce roughly
the same sound. This becomes obvious to anyone who tries out various
motor configurations in VTL and notices that there are many different
ways of producing, for example, the vowel $/i/$. In deciding whether
a given articulator representation of a speech sound is a good representation,
the following question is crucial: 

Is the articulator shape realistic (e.g. does the tongue seem naturally
positioned)?

In order to ensure this was the case in all my prototype vowels, I
checked with the preconfigured vowel configurations of the infant
and adult VTL speaker and visually compared the two shapes.

VocalTractLab was used throughout this thesis in two ways:
\begin{itemize}
\item Speech samples (ambient speech) were synthesised using VTL. These
are labeled and used for training the auditory system of the infant
learner. Ambient speech is produced by gaussian sampling near motor
parameters of the prototype vowels. Some of these samples will sound
like good representations of the prototype (e.g. prototype vowel $/a/$
of male speaker 4 (age 4)). These are given the label $/a/$. Some
of the samples (especially those farther from the mean), will not
really sound like a German vowel and are given the label $null$.
The Auditory System is then trained with most of these samples (training
set), and tested for accuracy on the rest of the samples.
\item VTL is used when babbling (imitation stage). The RL-agent explores
a set of (or all of) the 16 motor parameters by choosing values for
each parameter. VTL then accepts those parameters and produces a speech
sound\footnote{The VTL-produced sound might also be silent. This is the case when
the articulators are positioned in a way that completely cuts off
airflow through the vocal tract.}. In this application, VTL serves as part of the environment for the
Reinforcement Learner.\newpage{}
\end{itemize}

\subsection{\label{subsec:Auditory-System}Auditory System}

The auditory system is the part of the model that accounts for perception
of sound. Just like an infant has to learn language perception, supervised
learning is applied in order to train a neural network to distinguish
incoming sounds as different phonetic groups (vowels, in this case).
To distinguish (in this context) means: ``to what confidence can
I classify a sound as belonging to each class?''. After training
and when confronted with a good representation of a vowel, the auditory
system should return a high confidence for that specific class and
low confidences for all the other classes.

Our auditory system works with three components that take us from
a sound input to class-confidences:
\begin{itemize}
\item The cochlea model simulates sound processing in the inner ear. A speech
signal is tonotopically transformed into nerve activations when receiving
(tonotopically organised) sound frequencies. We use the dual resonance
nonlinear (DRNL) filter model as described in~\cite{DRNL1}. It is
implemented in ``BRIAN hears''~\cite{key-2}, an extension of the
BRIAN neural network simulator for auditory processing~\cite{key-3}.
The model covers the range of 100 Hz to 8 kHz for input sound and
returns neural activation for 50 channels.
\item An \textit{Echo State Network }(ESN) is what lets us simulate auditory
memory~\cite{ESN-Scholarpedia}. A static\footnote{Here, 'static' means that when learning (after being randomly chosen)
the weights in between neurons in the reservoir are kept constant.
Usually, neural networks learn all weights between neurons using (e.g.)
back propagation. Recurrent neural networks get rid of certain difficulties
of learning hidden-to-hidden connections by simply learning the linear
weights coupling neurons in the reservoir to the output neurons (class
neurons).} reservoir (a pool of recurrently connected neurons) connected to
the channels of the cochlea model. These neurons correspond to a simplified
auditory cortex. Neurons in the reservoir are connected to output
neurons (one for each class). The weights of these connections can
be learned by presenting correctly labeled samples to the ESN and
adjusting the connection weights.
\item Classification based on the ESN-readout. The readout is a set of neuron
activations over time (bounded by the duration of the speech sound)
for each class $v$ (vowel). After averaging each class's output activation
$a_{v}(t)$ over time $:=a_{v}$, a confidence for each class is computed
using the softmax function:
\[
c_{v}:=\frac{exp(a_{v})}{\sum_{i}exp(a_{i})}
\]
The confidence of one class acts as reward for the reinforcement learner
during the imitation phase.
\end{itemize}
Keep in mind that the auditory system (in particular the ESN) has
first to be trained to classify correctly (target acquisition). In
the imitation phase we then use the pretrained ESN to return rewards
to the reinforcement learner. The RL-agent targets a specific vowel
to learn \textendash{} the confidence of that specific target $c_{target}$
is then used as reward\footnote{To be precise: The reward is computed using the confidence. Other
factors like metabolic cost and penalties from overstepping boundaries
of the parameter-hypercube are also taken into account. More on this
in section \ref{subsec:RL-algorithm}.}.

\begin{figure}
\includegraphics[width=1\columnwidth]{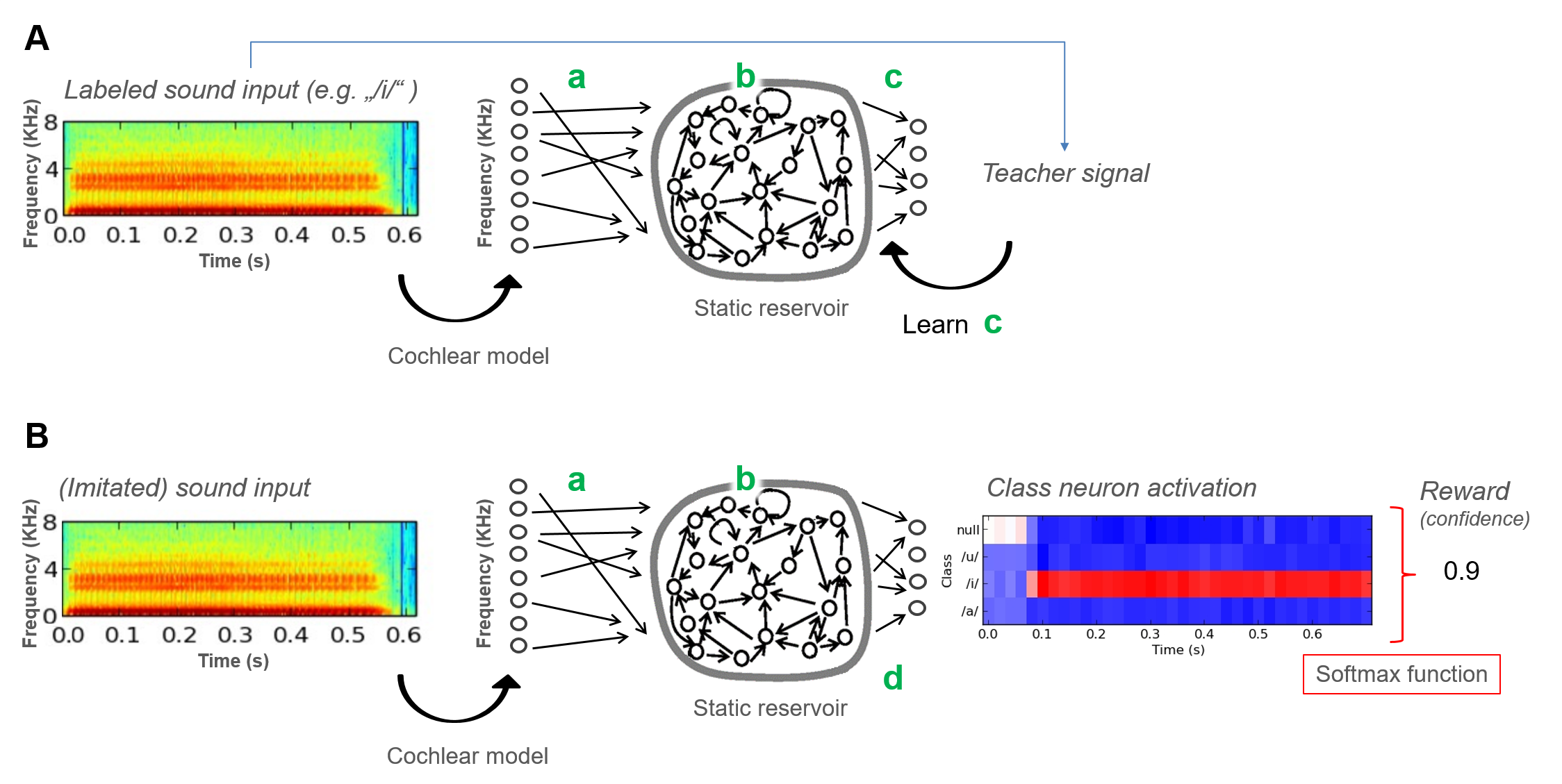}

\caption{\label{fig:Auditory System}The auditory system in target acquisition
(perceptual training) (\textbf{\textit{A}}\textit{).} During imitation
(\textbf{\textit{B}}), the auditory system is used as part of the
environment of the RL-agent, returning rewards for speech input produced
by the learner.\protect \\
The cochlear model transforms sound into nerve signals of tonotopically
ordered neurons (as in the cochlea). These are connected to the static
reservoirs using randomly set, constant connections (\textbf{\textit{\textcolor{teal}{a}}}).
Stochastic weights between neurons in the reservoir (\textbf{\textit{\textcolor{teal}{b}}})
are also kept constant after initialisation. Reservoir-to-output neuron
connections (\textbf{\textit{\textcolor{teal}{c}}}) are learned (linear
regression weights of the teacher outputs on the reservoir neuron
states) and then kept constant during imitation (\textbf{\textit{\textcolor{teal}{d}}}).}
\end{figure}

Both stages are illustrated, shown with the components of the auditory
system in Figure~\ref{fig:Auditory System}. All connection weights
that are randomly initialised are done so in a very specific manner.
Connections between neurons in the ESN are set in order that input
(e.g. speech input) 'echoes' around the network for some time. These
connections are very sparse (most weights are 0) and highly recurrent
(many connected neurons are connected in both directions). This creates
lots of loosely connected coupled oscillators. These oscillators (the
neurons of which they consist) are coupled to the output class neurons
which are learned during the perceptual stage. When imitating, all
connections are kept constant (the ESN has already learned to classify).

ESNs can, in principle, use any neuron model. In \textit{Listen and
Babble}, non-spiking leaky integrator neurons were used (we can switch
from leaky to non-leaky). 

One important parameter of an ESN is the size of the reservoir $N_{res}$
(the total number of reservoir neurons). The classification task is
a bit like fitting a (non linear) function to the data (speech sounds).
Increasing the number of neurons is similar to increasing the rank
of a polinom fit \textendash{} data is fitted more accurately with
increasing rank (or $N_{res}$) but the danger of overfitting also
increases.

Another parameter worth mentioning here is the \textit{spectral radius}
$\rho(\boldsymbol{W})$ (the maximal absolute eigenvalue of the Matrix
$\boldsymbol{W}$containing all the reservoir connection weights).
As a rule of thumb, $\rho(\boldsymbol{W})$ ``should be greater in
tasks requiring longer memory of the input''~\cite{ESN_Practical}.
Varying this parameter was not part of my work but could, in future,
be part of increasing memory of the auditory system when looking at
production of syllables (and no longer only phonemes).

For more on the used ESN parameters, see section 4.2.1 in~\cite{Thesis Murakami}.

In section~\ref{subsec:VocalTractLab} I explained how I produced
ambient speech from a group of speakers of different ages. No longer
training the ESN on the two standard speakers only but using a wide
variety of speech sounds to train for each class lets the auditory
system generalise over speech sounds from various ages. The accuracies
of such ESNs, trained with many speakers and with 2 more classes ($/e/$
and $/o/$) added, are shown in section~\ref{subsec:Quality-AS}.

I only altered the size of the ESN reservoir in perceptual training.
All other ESN parameters were carried over from Murakami's thesis.\newpage{}

\subsection{\label{subsec:RL-algorithm}RL algorithm}

An agent learning all the 16 motor parameters mentioned in previous
section~\ref{subsec:VocalTractLab} is equivalent to the agent finding
a certain point in a 16-dimensional motor space. This needle-in-a-haystack-problem
is solved using \textit{Covariance Matrix Adaptation Evolution Strategy
}(CMA-ES).

Since my focus shifted towards the perceptual stage in this thesis,
I will only very briefly cover the RL-algorithm and point the reader
to an abundance of descriptions found in the Web (e.g.~\cite{CMAES}).

CMA-ES is a realisation of Reinforcement Learning. An agent interacts
with its environment and, based (in this case) only on returned rewards,
optimises its behaviour. In our form of imitation we are confronted
with a black-box optimisation problem. The learner knows nothing about
the physics (or phonetics) happening in his vocal tract. He simply
positions the articulators and receives a reward (``how well the
produced sound represented a certain target''). CMA-ES is an algorithm
formulated for such cases. Points in the parameter hypercube are sampled
using a gaussian distribution (the sampled points in each iteration
are called \textit{sample generations}). The new $x_{mean}$ is computed
as the weighted average of the $\mu$ best samples of the last generation.
As the algorithm thus finds and converges in local reward-optima,
speech sounds of the RL-agent more and more accurately represent the
teacher signals (or prototype vowels).

\paragraph{Target choice}

Deciding on which target to learn is an important part of learning
efficiently. Humans experience extrinsic motivation (i.e. direct rewards
or punishments from the environment) as well as intrinsic motivation
(not driven by immediately useful rewards, but rather other concepts
such as innate curiosity, or the desire to increase one's competence).
Murakami discusses intrinsic motivation in the framework of Listen
and Babble in his thesis~\cite{Thesis Murakami} (p. 21-22, 30-31). 

As in the original model, I arbitrarily set the target at the beginning.
However, unlike Murakami, I did not finish learning a target until
choosing the next one. The original Listen and Babble model, after
achieving satisfactory rewards for a target (i.e. vowel), chose the
next target based resampling and choosing the highest confidence target
that was not yet learned. Thus, easier targets were learned first
until the RL-agent committed himself to targets that are harder to
reach. This reflects ideas on intrinsic motivation, where the infant
babbler could be motivated by making rapid progress, and leaving slower
learning until later, when easier targets have already been mastered.

I made a slight modification in that I let the RL-agent swap targets
in each generation (after receiving a reward for the generation of
samples and choosing the $x_{mean}$ for the next generation). If
confidences towards other targets are higher, the RL-agent chooses
the target with the highest confidence as his next target. ``I am
trying to imitate $/a/$. But what I just tried sounded more like
$/i/$. I'll take that as my target instead!''. This way the agent
is more directly driven by intrinsic motivation.

\cleardoublepage{}

\section{\label{sec:Results}Results}

\textcompwordmark{}

\subsection{\label{subsec:A-series-of}Ambient speech}

\subsubsection{Age-specific pitch}

Pitch, a prominent feature of speech, changes from infancy to adulthood.
I modeled this feature by performing spline interpolation on data
from~\cite{Lee_Formants,F01}. The actual values used for the speakers
are documented in the control parameters script (\textit{control/get\_params.py})~\cite{Repository}
and plotted in Figure~\ref{fig:Speaker_Series}~a.

\subsubsection{Speaker series}

\begin{figure}
\begin{centering}
\includegraphics[width=0.8\columnwidth]{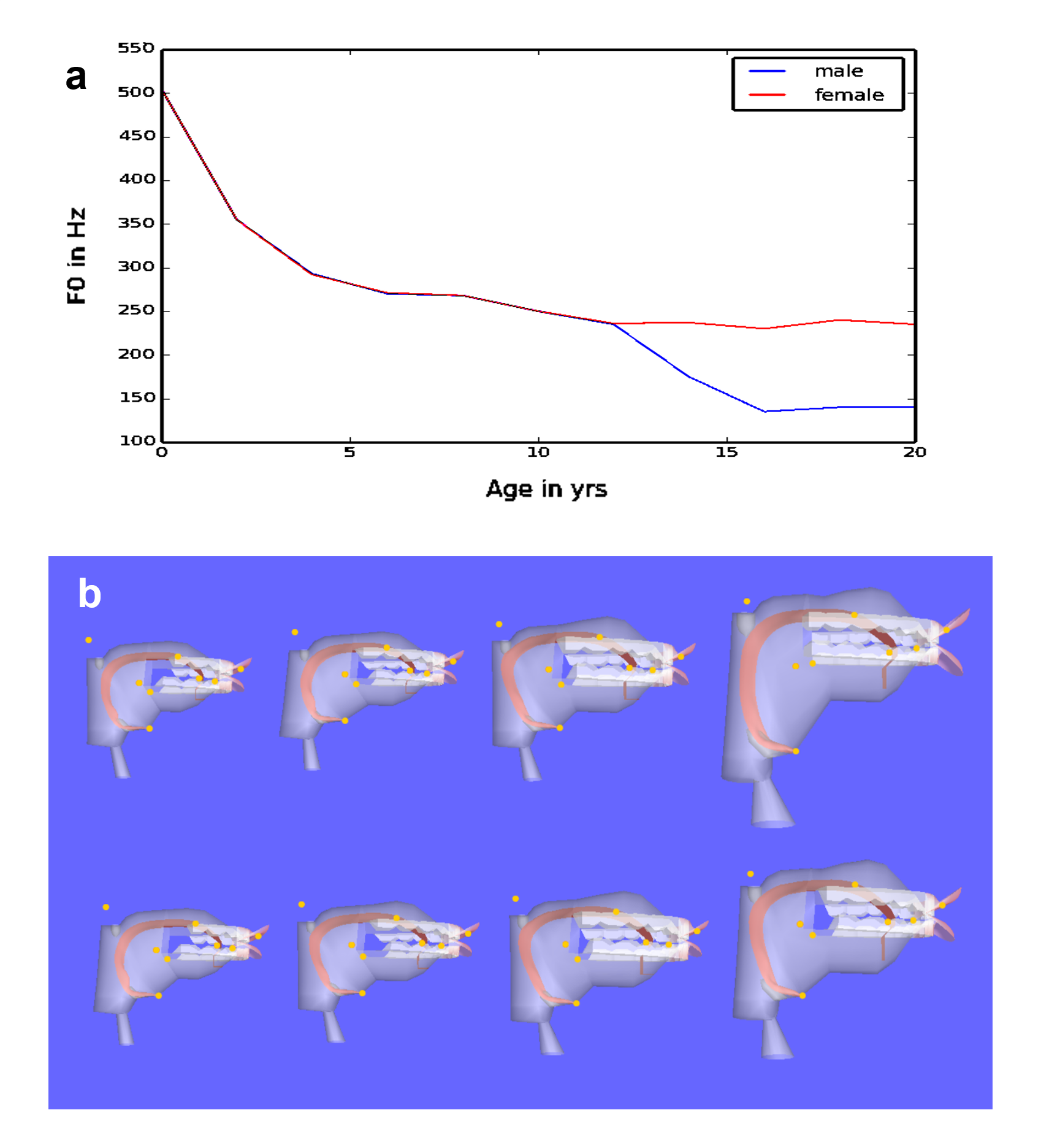}\caption{\label{fig:Speaker_Series}\textbf{a}: Pitch values used in the model
(both male and female) from 0 to 20 yrs (same age development as in
b). \textbf{b}: Selected speakers from the speaker series shown in
the VTL graphical interface. Top row from left to right: male speakers
aged 0, 4, 10, 20. Lower row from left to right: female speakers,
same ages.}
\par\end{centering}
\end{figure}

Simulating ambient speech for speakers of various ages meant producing
many different speaker anatomies. This speaker series was constructed
using VTL. Each speaker's anatomy was calibrated based on a model
of anatomical development implemented in VTL. I chose an age difference
of 2 years from one speaker to the next. This lets each speaker sound
audibly different to the next speaker in the series. Both sexes (22
speakers in total) were constructed using the following ages (in years):
\begin{flushleft}
\[
\left\{ \,0\,(newborn)\,,\,2\,,\,4\,,\,6\,,\,8\,,\,10\,,\,12\,,\,14\,,\,16\,,\,18\,,\,20\,\right\} .
\]
My speaker series differed from the anatomical model development in
VTL in the following points.
\par\end{flushleft}
\begin{itemize}
\item Overall size of the glottis (male/female) was derived from data given
in~\cite{anatomy1}.
\item Size of the vocal folds (male/female) is based on data given in~\cite{anatomy2}.
\item In the six-year-old male speaker, upper molar height was changed from
0~cm (VTL anatomy model) to 0.36~cm in order to give a smoother
transition between the 6-year-olds and the 8-year-olds.
\end{itemize}
For a 3-d visualisation of some of my speakers, see Figure~\ref{fig:Speaker_Series}~b.

\subsubsection{\label{subsec:Vowel-shape-settings}Vowel shape settings}

As a second step\footnote{These steps are shown in logical progression. In reality, multiple
speaker series were created and altered a few times to make them more
realistic. For each of these, vowel gesture parameters were set and
reset in order to get rid of certain disturbing effects in the audio
data (e.g. when airstream was too narrowly restricted, this could
produce a scratching noise in the actual data used as ambient speech).}, vowel gestures were modeled for each speaker. Since each speaker
has his own specific anatomy, the set of parameters used for a specific
vowel will look different from the previous speaker (or, in general,
all other speakers with slightly different ages). Modeling vowel gestures
resulted in a set of prototype vowels for each speaker. In this thesis,
I calibrated prototype articulator positions for the following vowels:
\[
/a/,/e/,/i/,/o/,/u/.
\]

Vowel shapes were set by hand in VTL. After listening to the resulting
(synthesised) speech sound of a hand set shape, I corrected the articulators
and again listened to the sound. This was repeated until I perceived
the speech sound to be a good representation of a given vowel.

I already mentioned in section \ref{subsec:VocalTractLab} that often
multiple articulator configurations can produce the same speech sound.
It is possible to produce all vowels in VTL with unrealistic positions
(e.g. of the tongue). A human calibrating a vocal tract in VTL needs
to be guided by knowledge of how vowels are indeed produced by humans.
I therefore invested some effort to become familiar with the articulators'
positions of the standard speakers in VTL for each vowel, since these
were set by a trained phoneticist.

I note here that neutral articulator positions ($/@/$ in SAMPA\footnote{\textit{Speech Assessment Methods Phonetic Alphabet} (SAMPA) is based
on the \textit{International Phonetic Alphabet} (IPA), the main difference
being that SAMPA uses ASCII characters. The schwa sound, is written
as $[@]$in SAMPA, but $[\text{\textschwa}]$ in IPA notation.} notation, also called \textbf{\footnotesize{}schwa} as in German
'bitt\textbf{\textit{e}}') are preset in the speaker anatomy model
of VTL. We therefore do plot $/@/$ , but remember that it is not
learned by the infant in this thesis, nor is it a distinct class in
the auditory training. $/@/$ positions are actually used as initial
articulator position for the RL-agent, from which the RL-agent explores
and learns $/a/,/e/,/i/,/o/$ and $/u/$. 

\paragraph{Prototypes}

In section \ref{subsec:VocalTractLab} I described how vowel samples
are generated for the perceptual learning. Using articulator positions
of the prototype vowels, these are slightly varied in order to produce
many prototype-like vowel samples. Sometimes, even the small changes
in the articulator positions mean a large change in how I perceived
the speech sound. This meant I had to label each sample before training
reservoirs (section \ref{subsec:Auditory-System}) on the data.

\begin{figure}
\centering{}\includegraphics[width=0.7\columnwidth]{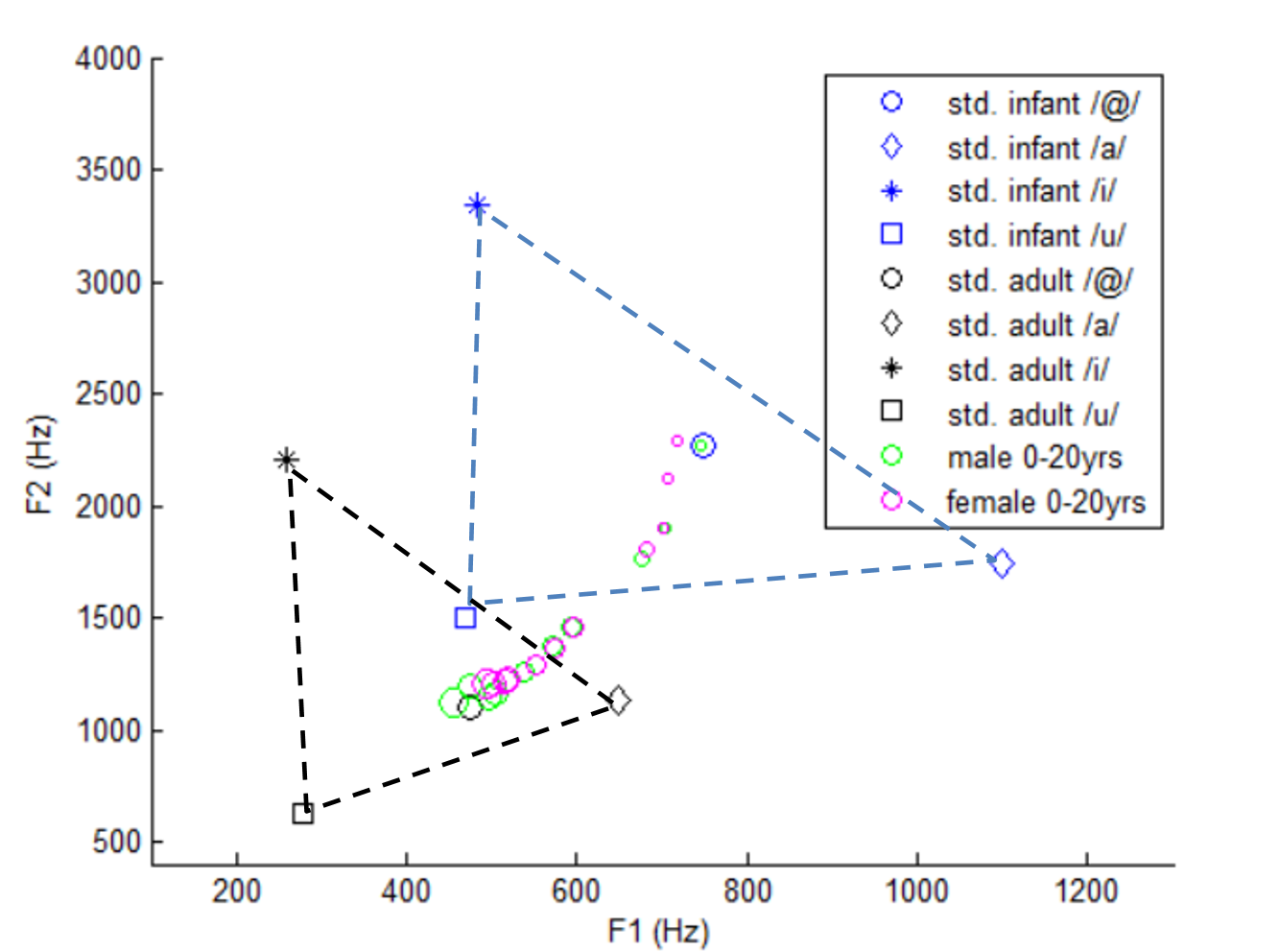}\caption{\label{fig:Vowel_Triangle}Vowel ($/a/,/i/,/u/,/@/$) formants of
the VTL infant speaker (blue), and adult (black). Green and pink cirles
are neutral articulator positions of my speakers, which form a trajectory
from infant $/@/$ to adult $/@/$. Larger circles stand for older
speakers. Lines drawn in between vowels serve only to illustrate (vowel
triangles).}
\end{figure}

In the original model, vowel samples near preset vowel shapes (of
adult and infant speakers) were used. The formants of those standard
vowels ($/a/,/i/,/u/$) spanned triangles in formant space, the adult
speaker's triangle having lower $F_{1}$ and $F_{2}$ values than
the infant speaker's vowel triangle (see Figure~\ref{fig:Vowel_Triangle}).
Since the first two formants $F_{1}$ and $F_{2}$ are central in
vowel recognition we can actually think of speaker normalisation in
terms of mapping age-trajectories through formant space, or ``which
paths do vowels take in formant space with in- or decreasing age''?

In Figure~\ref{fig:Vowel_Triangle}, speech sound formants $F_{1}$
and $F_{2}$ are shown for:
\begin{itemize}
\item VTL-preset adult speaker vowels (black markers)
\item VTL-preset infant speaker vowels (blue markers)
\item my speaker series' neutral articulator ($/@/$) - sounds.
\end{itemize}
It is easy to see that the $/@/$ formants form a trajectory from
the infant's $/@/$ to that of the adult\footnote{This way of thinking about speaker generalisation assumes static speech
sounds. Formant values up to section \ref{subsec:Motherese} are computed
by taking the median formant for that vowel. However, while pronouncing
one vowel, formants vary over time. This is especially significant
in infant-directed speech (with large $F_{0}$ variation).

For now, we shall assume constant formants for the duration of a vowel
(by taking the median as representative), and take into account changes
in formants \textit{over age}.}.

If vowel prototypes $/a/,/e/,/i/,/o/$ and $/u/$ of my speakers also
follow similar trajectories as do $/@/$ sounds in Figure~\ref{fig:Vowel_Triangle},
I take them to convincingly represent a development from adult to
infant. Thus, when trained on, they ought to work towards normalising
vowel perception of the auditory system (bridging the age gap between
the infant and adult speaker in perception). An auditory system trained
on samples from my speaker series should then be able to correctly
perceive vowels from any age group.

\begin{figure}
\begin{centering}
\includegraphics[width=1\columnwidth]{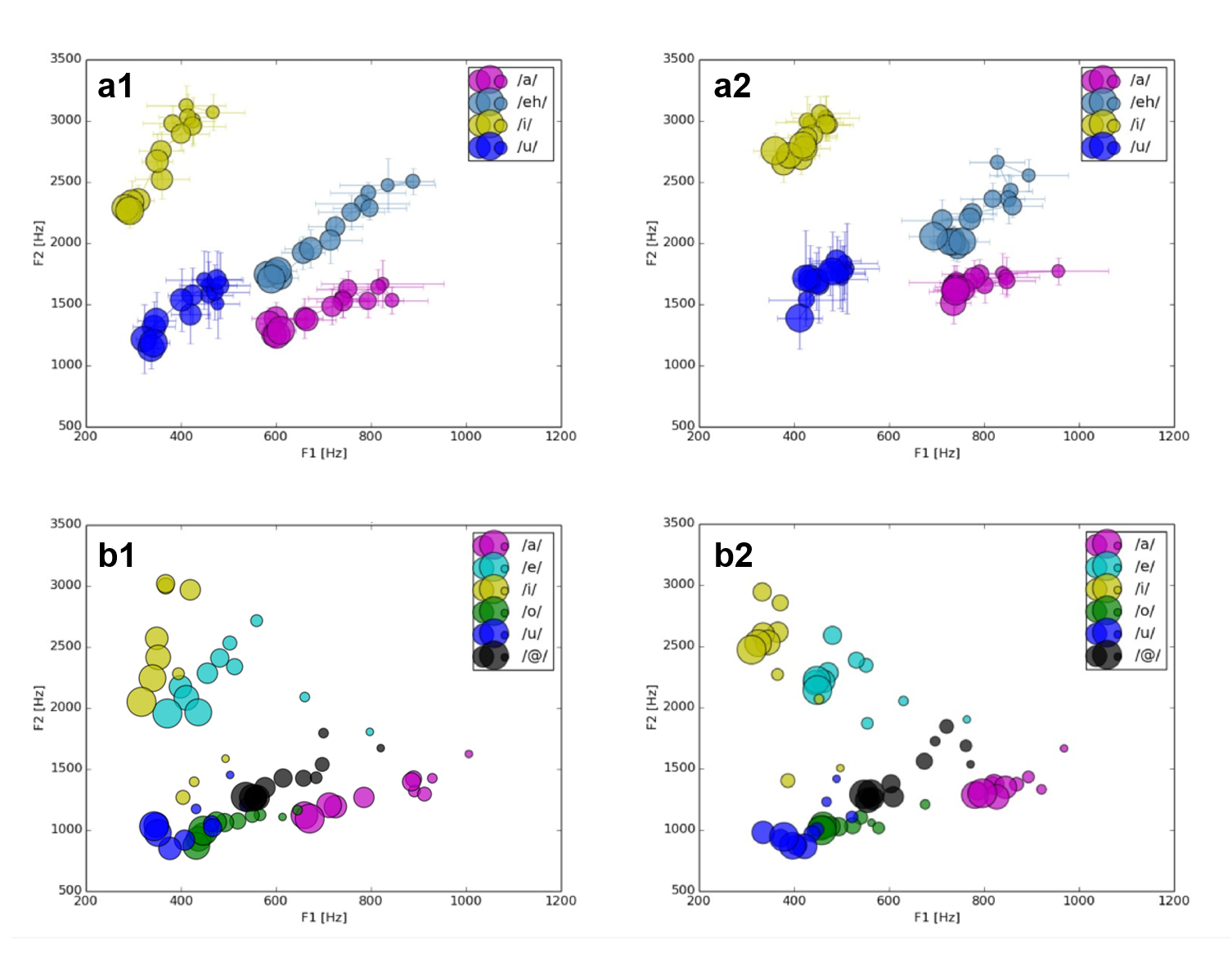}
\par\end{centering}
\caption{\label{fig:Proto_Formants_and_Literature}English vowel formants \cite{Lee_Formants}
(\textbf{a1}: male speakers and \textbf{a2}: female speakers) compared
to German vowel-prototype formants synthesised in VTL (\textbf{b1}:
male VTL speakers, \textbf{b2}: female VTL speakers). Literature formant
error bars are the standard deviations among different subjects' vowels
from the same age group. The size of the circle indicates the age
(the larger the circles, the older the speakers).}
\end{figure}

Figure~\ref{fig:Proto_Formants_M_F}~a~and~b show formants of
all speakers' vowels (male and female separate) including the neutral
position ($/@/$), and Figure~\ref{fig:Proto_Formants_M_F}~c shows
all the prototypes (only including vowels used in perceptual training,
without $/@/$). Some observations can be made:
\begin{itemize}
\item The vowel prototypes show a similar development as the neutral positions
do in formant space.
\item Some vowel groups strongly overlap (especially $/o/\leftrightarrow/u/$
).
\item Vowels from younger speakers (smaller cirlces) are less regular and
further apart.\medskip{}
\begin{figure}
\begin{centering}
\includegraphics[width=1\columnwidth]{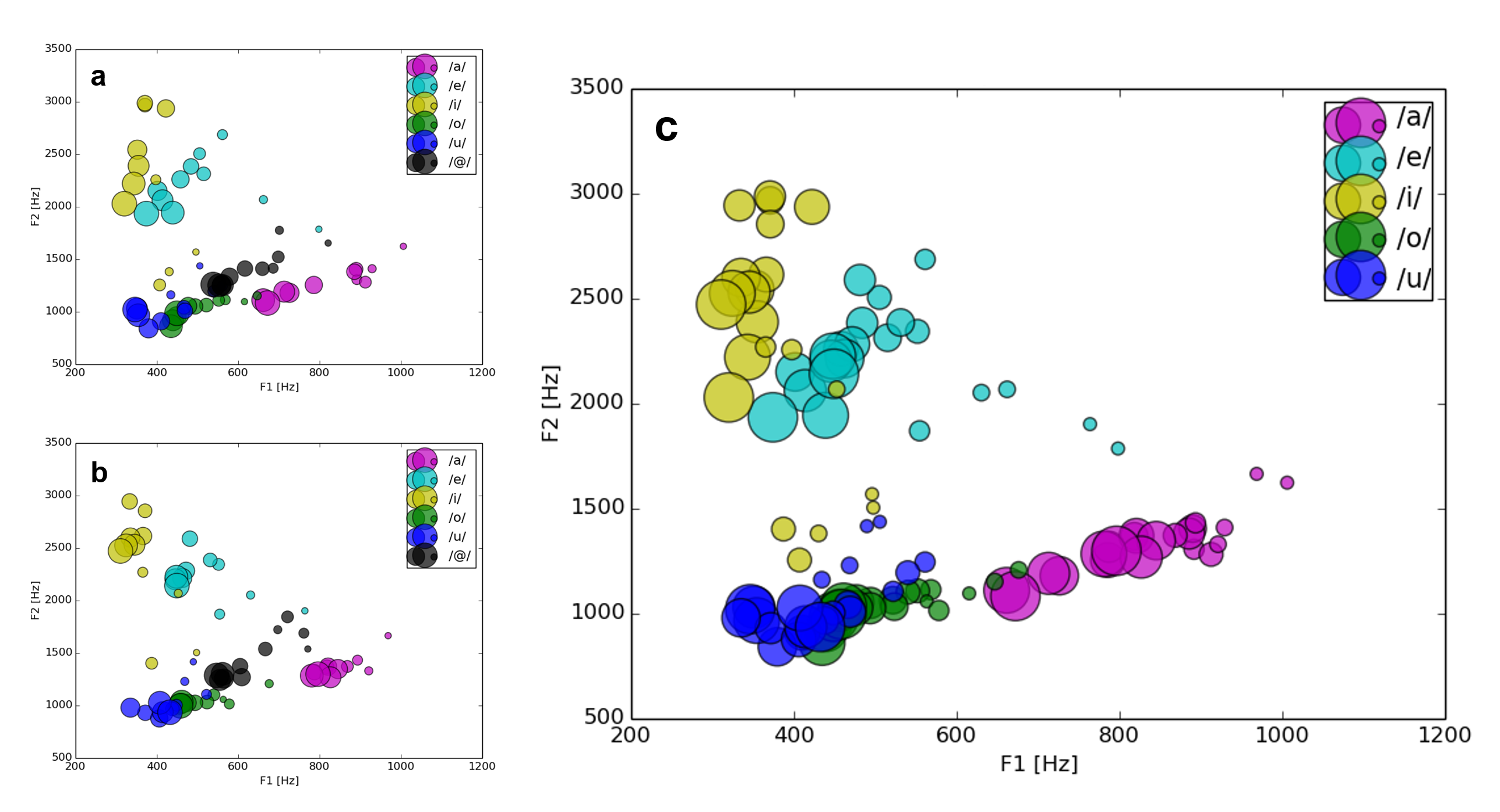}
\par\end{centering}
\caption{\label{fig:Proto_Formants_M_F}Formants of prototype vowels of my
speaker group. Larger circles stand for older speakers. Male speakers
(\textbf{a}) and female speakers (\textbf{b}) shown separately, including
$/@/$. \textbf{c}: All prototypes.}
\end{figure}
\end{itemize}
Despite the fact that shapes weren't set by a trained phonetician,
vowel prototypes do resemble vowels recorded by humans of various
ages. Formants from most protoype vowels display systematic developments
over age, just as is the case in real speech. In Figure~\ref{fig:Proto_Formants_and_Literature},
a formant space plot is shown for the male and female speakers compared
to formants of General American vowels recorded from human speakers
of age 5 to 19+\footnote{The group 19+ representing a group of speakers up to the age of 50.
Speech development slows down above the age of 20, which justifies
such a category. I am not aware of phonetic data of vowels available
for children younger than five years (various studies show other phonetic
features down to the age of four).}~\cite{Lee_Formants}. Note that their formants show similarities,
although realisations of General American vowels $/a/,/eh/,/i/,/u/$
are phonetically slightly different to those of German vowels $/a/,/e/,/i/,/u/$
(shifted in formant space). The following three additional observations
can be made:
\begin{itemize}
\item Similar vowels (marked with the same colours) take up similar parts
of formant space.
\item We see a similar development from young speakers' vowels towards older
speakers' vowels.
\item Though younger speakers again seem less regular as to their formants\footnote{This attribute of the young (age 6 and under) speaker prototypes of
my speaker group is actually feasible. Lee et al. also find that younger
speakers have considerably more variation in their speech's phonetic
features. \cite{Lee_Formants}}, this is not seen as clearly in the literature data (with no data
for speakers under 5).\medskip{}
\end{itemize}

\subsubsection{Vowel Samples}

Using the shape parameters (which mark the articulator positions)
of the prototype vowels, I sampled near (gaussian sampling with $\sigma=0.01$)
those parameters in order to produce vowel-representative samples,
and further away ($\sigma=0.2$) in order to produce non-representative
speech sounds (sounding unlike any of the vowels). All these samples
were then listened to individually and marked as either $/a/,/e/,/i/,/o/,/u/$
or $null$. In order to minimise any biases while labeling, samples
were shuffled and then listened to blindly. Between the speech sounds
I listened to 1 second of gaussian noise in order to reduce perceptual
biases due to the phonetic context. 

\begin{figure}
\begin{centering}
\includegraphics[width=0.7\columnwidth]{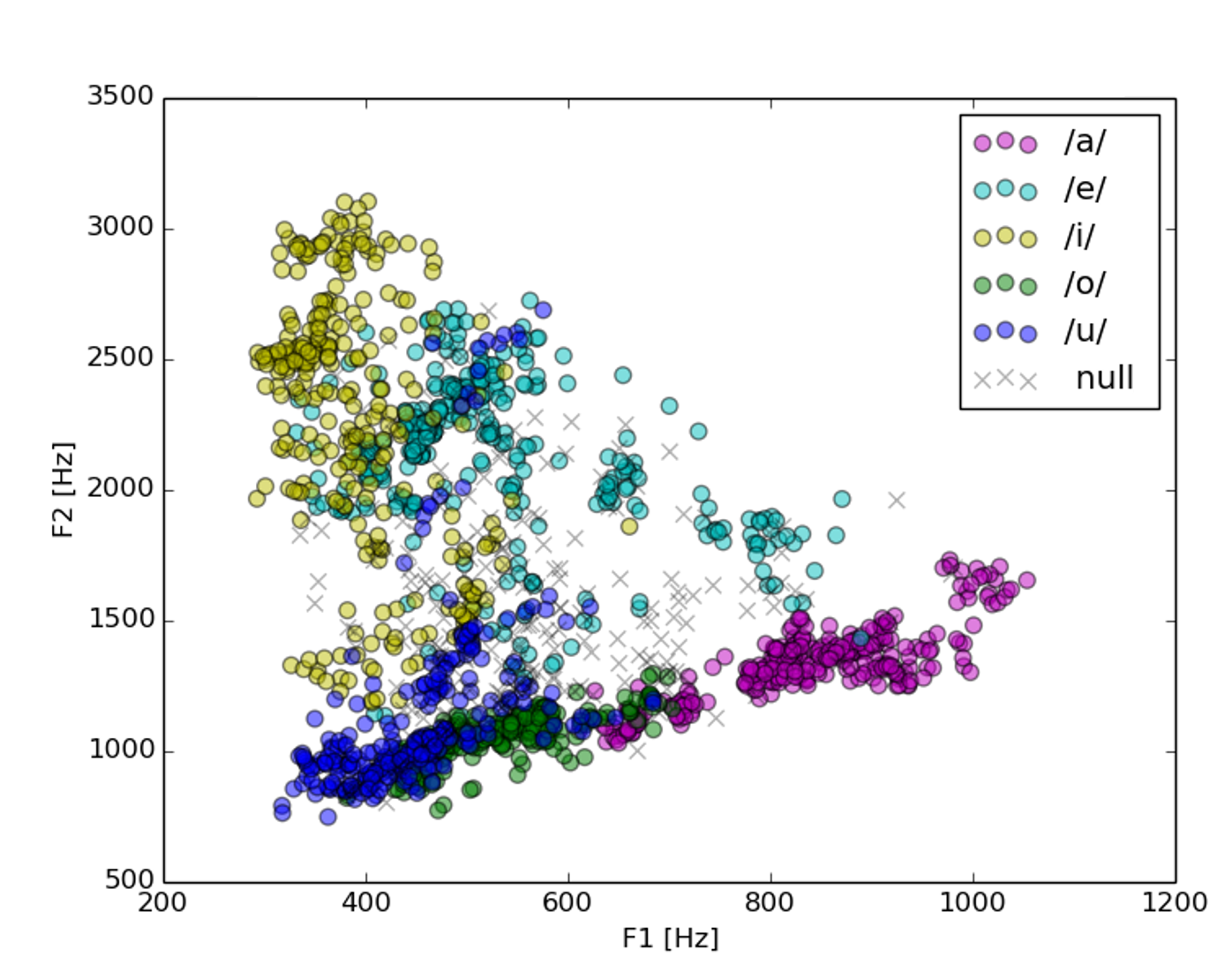}
\par\end{centering}
\caption{\label{fig:Vowel_Samples}Vowel samples in formant space. $null$
represents samples sounding nothing like any of the listed vowels.}

\end{figure}

Although sampled around prototype vowels, samples displayed much larger
formant variation. This is seen in Figure~\ref{fig:Vowel_Samples}.
We can again make some observations:
\begin{itemize}
\item Vowel groups, although sampled near prototype parameters, can be phonetically
quite dissimilar (compare with Figure~\ref{fig:Proto_Formants_M_F}~c).
\item Sample vowel groups (and thus, the classes which the auditory system
must learn to distinguish) strongly overlap in formant space (especially
groups $/o/$ and $/u/$, and $/e/$and $/i/$ respectively).
\item Of all groups, $/a/$ samples seem most organised.
\item Samples with formants far from the respective 'normal' area in formant
space were still recognised as belonging to that group.
\end{itemize}
\medskip{}

\newpage{}

\subsection{\label{subsec:Quality-AS}Training reservoirs}

The produced samples (including their labels) now act as ambient speech
for the (modeled) infant. A set of 2112 labeled samples are used in
the following steps (16 samples per speaker and per vowel). Auditory
systems with various reservoir sizes were trained on the ambient speech.
We now train with 6 distinct output classes: $/a/,/e/,/i/,/o/,/u/$
and $null$. 80\% of the samples were used to train each auditory
system with 20\% of samples used as a test set.

The reservoir is trained using the training set and then given the
task of classifying the samples in the test set. Rates of wrong classification
(error rates) are an easy measure of the accuracy of a specific reservoir.

\subsubsection{Partial training}

What happens if we only partially train a reservoir? In real life,
the infant learner should be able to solve the perceptual task without
hearing its own speech samples. Solving the problem of speaker generalisation
would mean being able to train a reservoir (for the auditory system
of the infant) with speech samples other than its own and then still
correctly being able to categorise its own speech sounds.

\begin{figure}
\begin{centering}
\includegraphics[width=0.7\columnwidth]{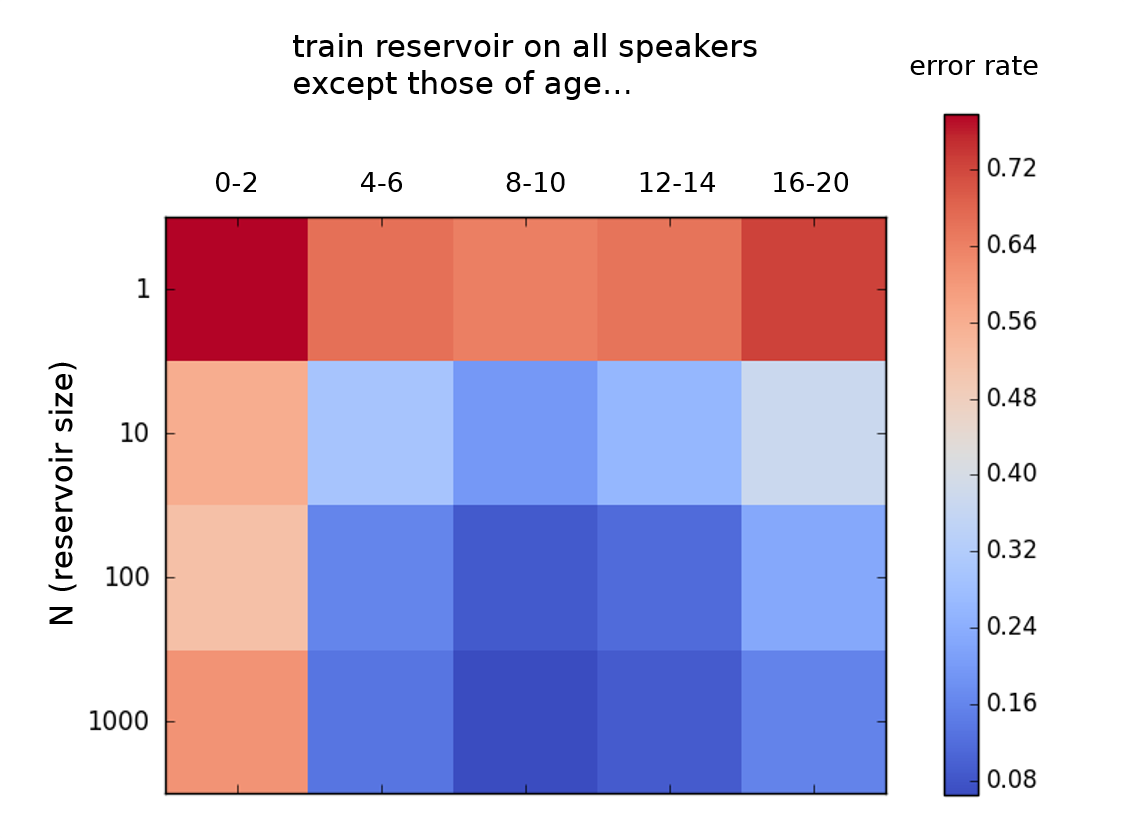}
\par\end{centering}
\caption{\label{fig:Partial_ESN_training}Mean error rates (misclassification
rates) over 30 trials for various reservoirs. ESNs of different reservoir
size $N$ were trained using training sets that omit samples from
a certain subgroup of speakers. The reservoirs are then tested on
that subgroup.}
\end{figure}

Normally when training a reservoir, test and training set are randomly
picked from the entire data set. But what if we remove all samples
produced by speakers of ages 0\textendash 2~yrs when learning, and
then use those samples (0\textendash 2~yrs) when testing the reservoir?
Figure~\ref{fig:Partial_ESN_training} shows error rates for 20 different
reservoir training paradigms, exploring the ability to generalise
for specific age groups. Reservoirs with sizes $N=\{..,10,100,1000\}$
were trained. For each reservoir size (e.g. $N=1$, first row in the
plot) errors are plotted (colour) for five different training paradigms.
The far left column, for example, shows reservoirs trained with samples
from speakers of all ages except 0\textendash 2~yrs. Samples from
speakers of 0\textendash 2~yrs served as test set. Error rates are
the mean error rates taken from 30 individual trials. While all subgroups
are made up of 4 speakers (2 male and 2 female), the last group (age
16-20) has 6 speakers. \\
We can observe that:
\begin{itemize}
\item Errors are smallest in the center column (omitting speakers aged 8-10).
Interpolating thus seems easier than extrapolating (perceiving speech
sounds from very young or very old speakers when not being trained
on these).
\item This extrapolation error (or: inability to normalise) is most strong
for young speakers, who display larger phonetic variance in vowel
production.
\item While on the whole error rates decrease for larger reservoir sizes,
this is not the case when extrapolating to young speakers (first column)
\textendash{} the $N=100$ reservoirs, trained without ages 0\textendash 2~yrs,
show smaller error rates then the $N=1000$ reservoirs trained without
the same subgroup. The larger reservoirs seem to overfit the data
in this case.
\end{itemize}
\medskip{}

\newpage{}

\subsubsection{\label{subsec:Comparing-reservoir-training}Comparing reservoir training
paradigms}

\begin{figure}
\begin{centering}
\includegraphics[width=1\columnwidth]{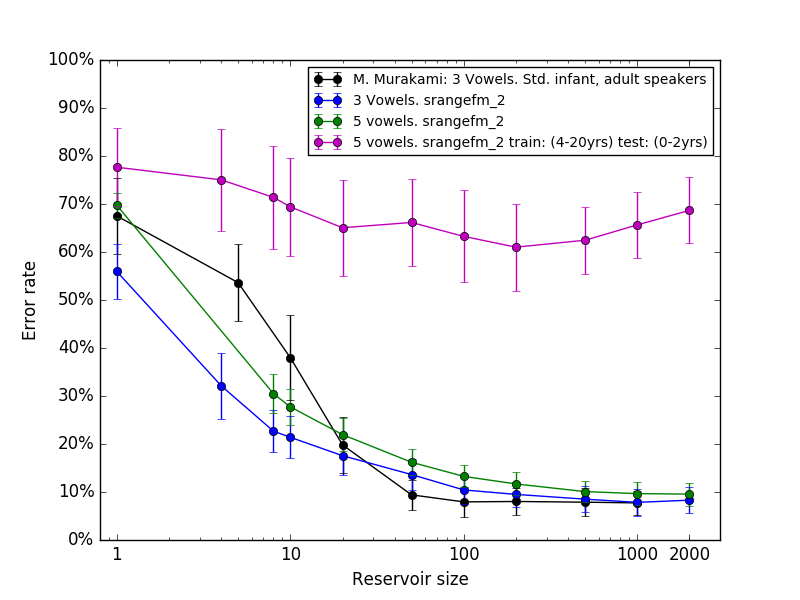}
\par\end{centering}
\caption{\label{fig:Error rates}All curves show error rates (rates of misclassification)
of the samples' test set for different reservoir sizes. Curve colours
code for different training (and testing) configurations.\textit{}\protect \\
\textit{Black:} Reservoirs trained on adult and infant speaker samples
and four classes ($/a/,/i/,/u/,null$ ) from \cite{Thesis Murakami}.\protect \\
\textit{Blue:} Reservoirs trained on four classes, but using samples
from the whole speaker range (0-20yrs, male/female).\protect \\
\textit{Green:} Reservoirs now trained on six classes (new vowels
$/o/$ and $/e/$ added), using samples from all speakers.\protect \\
\textit{Red:} Reservoirs trained on six classes, on samples from all
speakers that are 4 yrs and older (leaving out 0\textendash 2~yrs)
and testing on speakers of age 0\textendash 2~yrs. (Extrapolated)\protect \\
Each data point is the mean error rate of 100 trained ESNs; the error
bars are the corresponding standard deviations.}
\end{figure}

In order to make meaningful conclusions on how ambient speech from
a diverse group of speakers affects the quality of the auditory system
classification, I compare four different reservoir training paradigms.
Figure~\ref{fig:Error rates} shows rates of misclassification plotted
over different reservoir sizes. Training paradigms were the following:
\begin{enumerate}
\item Reservoirs are trained on only the standard infant and standard adult
speaker, discerning four classes: $/a/,/i/,/u/,null$ (\textit{black}). 
\item Reservoirs are trained on the whole speaker group, discerning the
same four classes (\textit{blue}).
\item Reservoirs are trained on the whole speaker group, now discerning
six classes ($/e/$ and $/o/$ included) (\textit{green}). 
\item Reservoirs try to extrapolate. Trained on all speakers that are four
yrs and older (leaving out 0\textendash 2~yrs) but testing on speakers
of age 0\textendash 2~yrs. These reservoirs attempt a form of speaker
normalisation. (Trained on all six classes) (\textit{red}). 
\end{enumerate}
Error rates from paradigm one were taken from Murakami's thesis~\cite{Thesis Murakami}.

\pagebreak{}Observations:
\begin{itemize}
\item The increased speech variety of a speaker group (compared with only
two speakers) actually makes correct classification easier for smaller
reservoirs. However, reservoirs of $N>50$ start to efficiently fit
data from only two speakers. Differences vanish for larger reservoirs
$N>500$ (compare \textit{black} with \textit{blue} curve).
\item When discerning two more classes we do not see a significant difference
in error rate (compare \textit{blue} with \textit{green} curve). We
must note that the chance-classification error rate would also be
slightly higher for six classes (83\%) than for only four classes
(75\%). Thus, the error rates per class would not be significantly
different when comparing paradigm 2 and 3.
\item Reservoirs have difficulties classifying correctly when extrapolating
towards young speakers (\textit{red (magenta)} curve). Very large
reservoirs seem to overfit the data.
\end{itemize}
\medskip{}

\begin{figure}
\begin{centering}
\includegraphics[width=1\columnwidth]{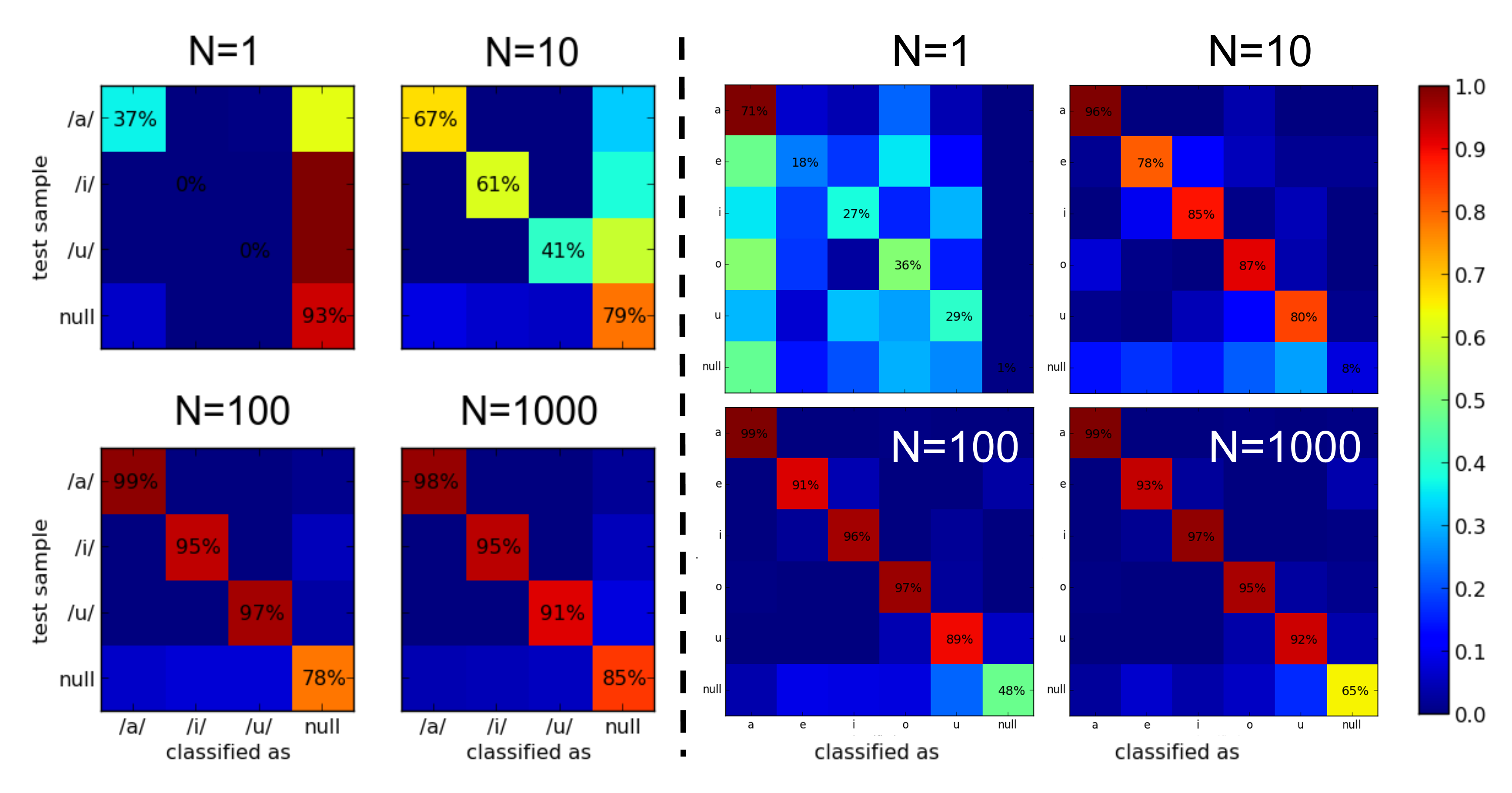}\caption{\label{fig:confidences}Confusion matrices for Murakami's thesis (\textit{left})
and for this thesis (\textit{right}). When a reservoir is tested for
its accuracy, a confusion matrix shows how many samples from the test
set (vertical axis) are correctly classified as their original labels
(horiz. axis). N denotes reservoir size.}
\par\end{centering}
\end{figure}

In Figure~\ref{fig:confidences}, I compare reservoir training paradigm
1 and 3. Despite having to classify more vowels and from a larger
variety of speakers, vowels are classified with a correct classification
rate that is comparable to Murakami's reservoirs. The main difference
is the $null$ class: Samples which were labeled as $null$ by me
were frequently (in 35\% of all cases!) classified as one of the vowels
by the reservoirs. The reservoirs are ``too tolerant'' about speech
sounds; bad quality vowels are given higher confidence levels than
they should be given. Also, vowel groups that strongly overlap in
formant space (see, for example, Figure~\ref{fig:Proto_Formants_M_F})
are more frequently mixed up (groups $/e/,/i/$ and $/o/,/u/$).

\medskip{}

\newpage{}

\subsection{Imitation}

Due to time limitations, I do not have statistically relevant results
as to how well the imitating RL-agent performs on a generalised auditory
system (as in section~\ref{subsec:Comparing-reservoir-training},
paradigm 3). However, I will show the result of one simulation using
a reservoir of size $N=100$, trained on all speech samples from my
speaker group.\\
\begin{figure}
\centering{}\includegraphics[height=0.7\textheight]{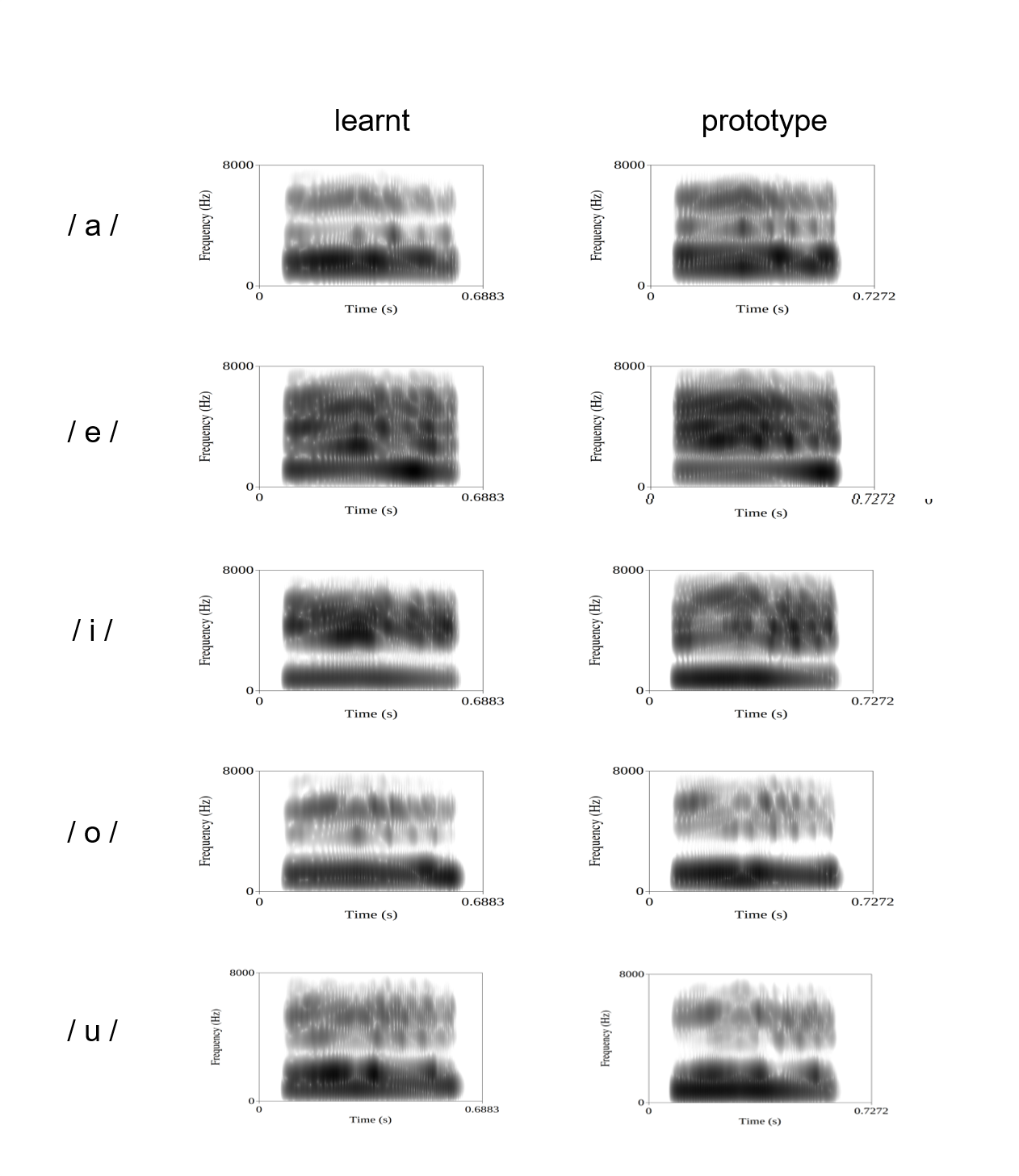}\caption{\label{fig:Spectro_Comparison_Learn}Spectrograms of learned vowels
(maximal reward in 1000 iterations of 10 samples) compared with prototypical
vowels from the VTL infant speaker as learner.}
\end{figure}

The RL-agent was the VTL infant speaker, whose speech samples are
not included in the reservoir training. The agent learned 13 motor
parameters, using mentor lip positions and jaw opening parameters.
This corresponds to the case of visually guided learning (see \cite{Thesis Murakami},
section 2.2.3).

\begin{figure}
\begin{centering}
\includegraphics[width=1\columnwidth]{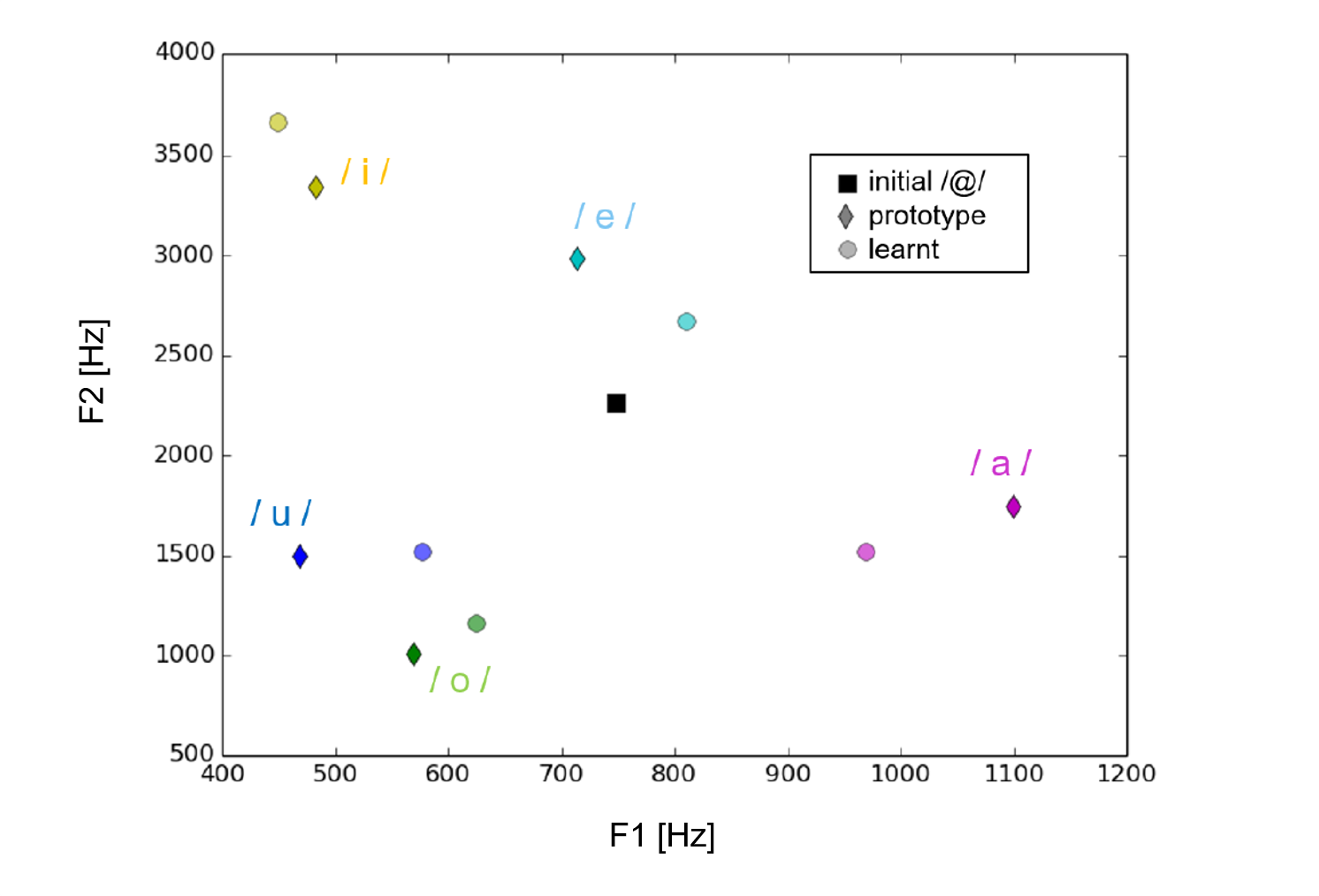}
\par\end{centering}
\caption{\label{fig:Formant_space_Learning}The learner no longer has one perceptual
target, but is trained on speech from all different ages. Learned
vowels in formant space (weakly) resemble prototypical vowels (preset
in VTL). Learned vowel space is smaller than prototypical vowel space
(nearly all learned vowels are closer together).}

\end{figure}

I did not use the FIAS cluster and did not yet implement the algorithm
in a parallel fashion\footnote{This is easily done. A proposed implementation of the RL-learning
is noted in the documentation and in the source code \cite{Repository}.}. Results are shown for $10,000$ samples (1000 generations of 10
samples). The RL-agent managed to approximate the target vowels so
that they were easily (though not perfectly) recognisable as that
vowel when I listened to them. Results are shown for the parameters
yielding the best rewards from among all the samples (not nescessarily
the configuration at the last generation!). In Figure~\ref{fig:Spectro_Comparison_Learn},
learned vowel spectrograms are compared to prototype vowels. In Figure~\ref{fig:Formant_space_Learning},
formants $F_{1}$ and $F_{2}$ are plotted for prototype vowels and
learned vowels. Unlike in Murakami's thesis, the prototype vowels
do not act as direct targets or mentor configurations. Murakami used
samples near the articulatory configurations of mentor vowels of the
infant speaker himself (along with those of the adult speaker) for
the perceptual training. I did not use samples from the (RL-agent)
infant speaker. The auditory system is trained on samples from other
speakers (Figure~\ref{fig:Vowel_Samples} showed their samples' distribution
in formant space).

\medskip{}

We can observe that learned vowels span a smaller vowel space than
hand-set vowel prototypes. The babbler seems to be happy with approximating
the vowels. With the exception of $/i/$, vowels seem to be nearer
to the neutral position and thus not articulated as distinctly as
prototype vowels are.

Since CMA-ES learning is a stochastic process, this result should
be reproduced and statistically reliable data gathered (e.g. mean
learning time for each vowel to yield a certain confidence level).
I sometimes even achieved sucessful imitation for less than 1,000
samples per target (roughly 4,000 samples for all 5 targets).

Another observation is that learning seemed to work better when learning
more targets. Remember that the RL-agent is intrinsically motivated,
switching from one target to the next. This could mean that learning
$/o/$ could aid in learning (phonetically similar) $/u/$. This too
should be subject to further investigation.

\newpage{}

\subsection{\label{subsec:Motherese}Motherese}

As we saw in \ref{subsec:Infant-directed-speech}, motherese helps
infants solve the perceptual problems in language learning. Infant
directed speech could be integrated in \textit{Listen and Babble}
and might even help solving speaker normalisation.

I tried to include motherese-like pitch variation in VTL samples.
But labeling VTL-synthesised speech with pitch variation over one
or two semitones proves difficult: what sounds like a specific vowel
changes and seemes to better represent another when pitch (alone)
varies too much. This perceptual experience is confirmed by Miller
\cite{Miller F0 tests} and by Slawson \cite{Mi_ Slawson F0} in their
research. Miller, for instance, found vowel category boundary shifts
in formant space for most English vowels when doubling the pitch (one
octave). $F_{1}$ boundary can shift from 100~Hz to 200~Hz for a
similar change in pitch (200~Hz) \cite{M_ Fuji F0}.

I noticed this perceptual shift, especially when hand-fitting vowel
shapes in VTL for very young speakers ( $F_{0}$ between 400~Hz and
500~Hz). I included slight pitch variations in all samples (two semitones),
but noticed that vowels seemed to shift in the way I perceived them
(most often between /i/ and /e/ and between /o/ and /u/ or /@/).

When using high pitch modulation in speech we noticably constrict
our own vocal tract, especially in the area of the pharynx. We never
only change the pitch, but ``accompany'' pitch change with the whole
vocal tract in order to perceptually stay in the same vowel category.
This means that we would have to modulate not only pitch but also
some (if not all) articulators in VTL. Laying out trajectories for
articulator parameters over time in order to stay within the same
vowel category the entire duration of the speech gesture is not easily
done.

Instead, I pursued the possibility of using human motherese speech
samples to train the auditory system. My wife kindly provided me with
speech samples of German vowels built into the following linguistic
context:

\[
\lyxmathsym{\textquotedblleft}Baba-bebe-bibi-bobo-bubu!\lyxmathsym{\textquotedblright}
\]

\begin{figure}

\begin{centering}
\includegraphics[width=1\columnwidth]{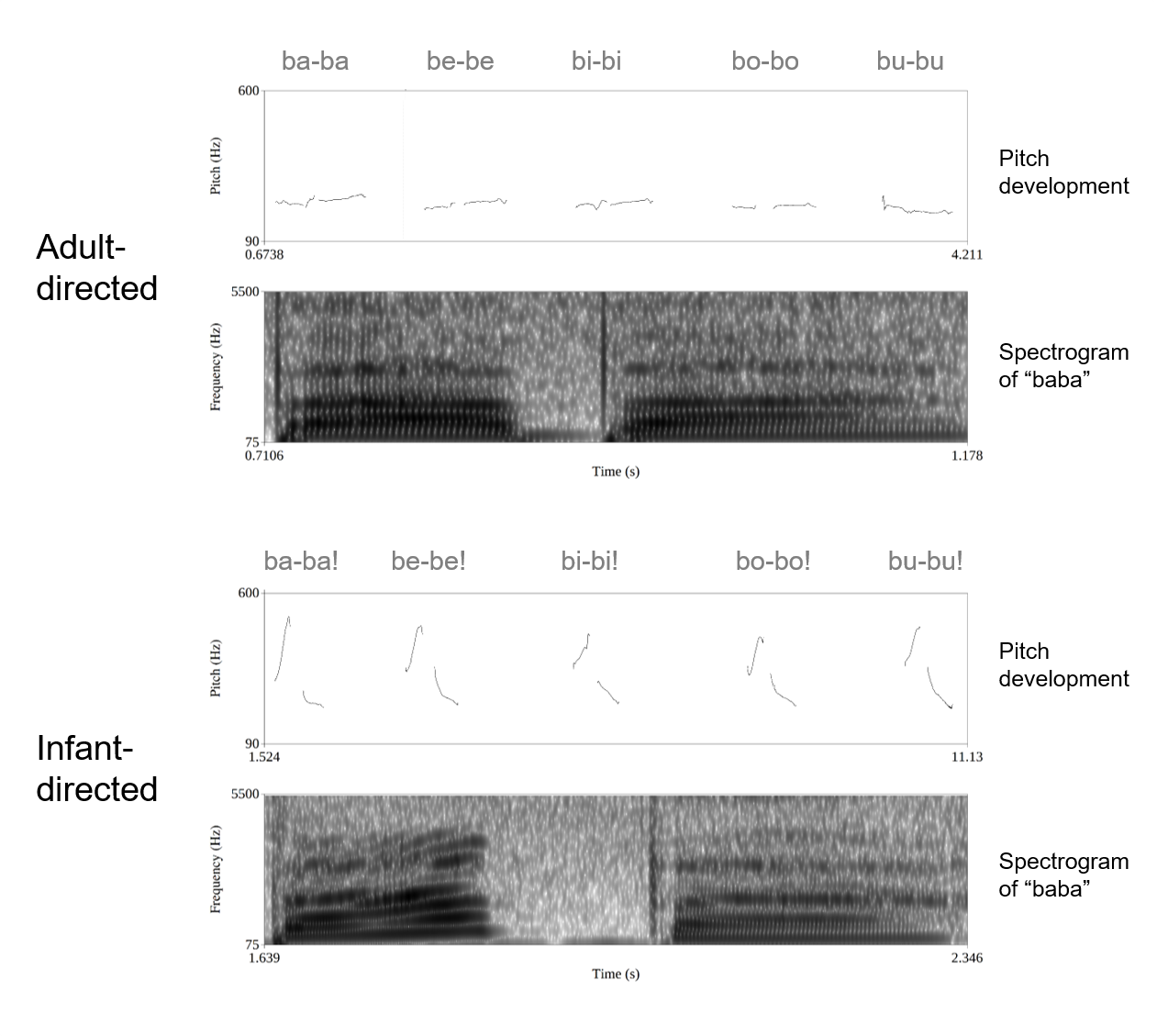}\caption{\label{fig:F0 Motherese Tirza}Pitch ($F_{0}$) variation in adult-directed
speech and in infant-directed (motherese) speech. Spectrograms shown
for the first ``word'' in the sentence.}
\par\end{centering}
\end{figure}

The sentence was first spoken in an adult directed way, then in an
infant-directed way. Figure \ref{fig:F0 Motherese Tirza} shows $F_{0}$
of the entire sentence and the spectrogram of the first ``baba''
in adult- and infant-directed speech. It is obvious that not only
$F_{0}$ but also the formants show large alteration throughout the
speech gesture in motherese speech.

\begin{figure}
\centering{}\includegraphics[width=1\columnwidth]{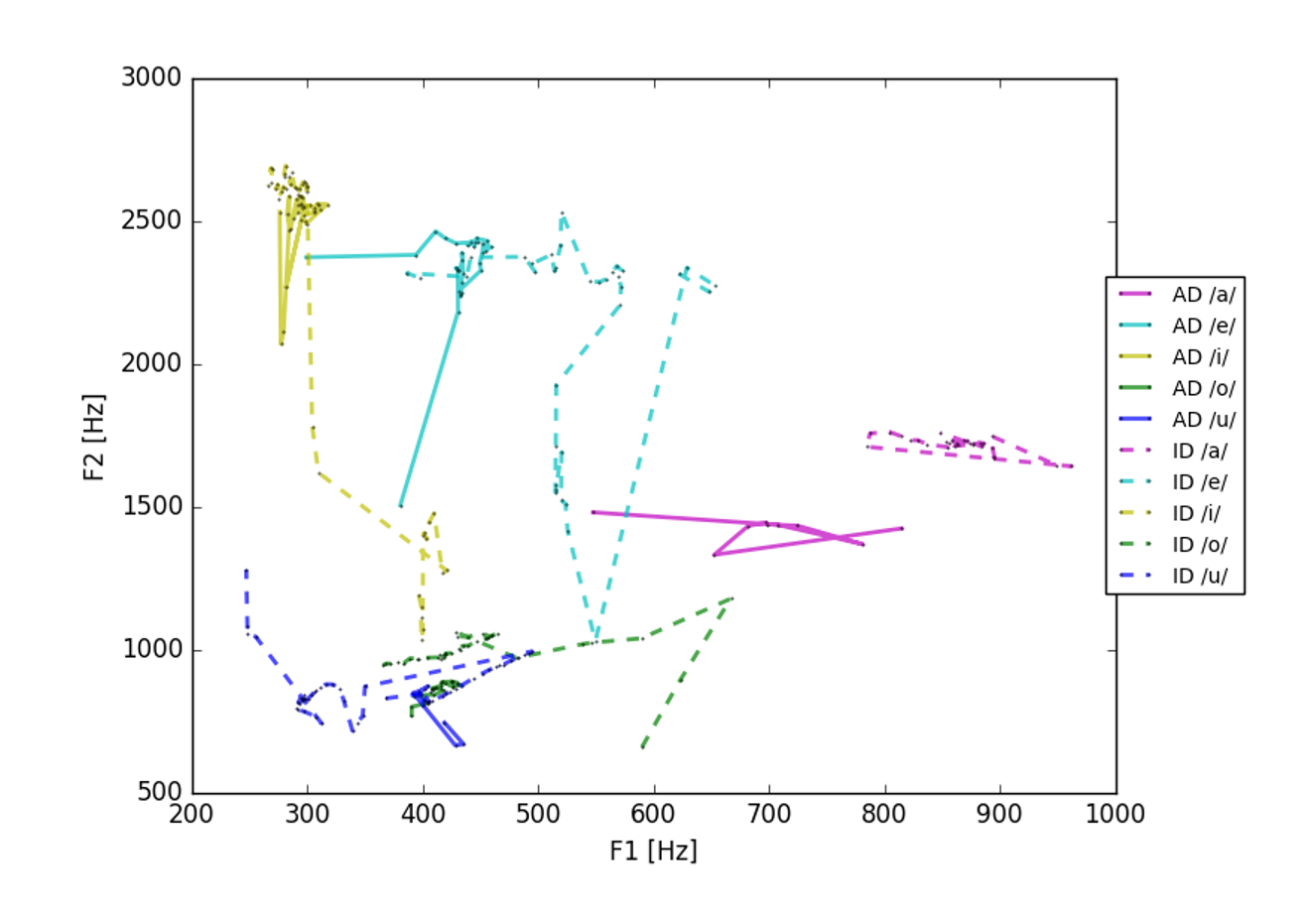}\caption{\label{fig:German-vowels-spoken}German vowels spoken infant-directed
(ID) and adult-directed (AD) in phonetic context. ID speech is marked
by larger formant space trajectories and larger vowel space. AD vowel
/o/ is hardly visible, and shows no formant development at all. Also,
in this case of adult directed speech /u/ and /o/ prove almost impossible
to discern using only the formants. Formants of $/o/$ (AD) are located
very near those of $/u/$(AD) and are not so clearly seen because
they are nearly static.}
\end{figure}

I found that extracting formants is not trivial in a linguistic context\footnote{Visible horizontal lines in the spectrogram are not necessarily the
actual precise formants!}. My own, male speech proved difficult to read formants from but I
managed to produce the trajectories for the female-spoken vowels.
The result was larger and more distinct formant trajectories in formant
space for infant directed speech, see Figure~\ref{fig:German-vowels-spoken}.

\cleardoublepage{}

\section{\label{sec:Discussion}Discussion}

\textcompwordmark{}

In this section, I discuss what I think my work demonstrates, and
also in which direction I believe my results point, regarding future
work.

\paragraph{Implementation of \textit{Listen and Babble}}

The model went through a large amount of changes concerning implementation
(code). The project structure should enable more straightforward ways
of introducing new steps (like those I will cover in this section).
For a more detailed understanding of how the code works, I refer to
the documentation on the \textit{github repository~}\cite{Repository}.

\paragraph{Speaker normalisation}

The thrust of my research shows: although using a range of speakers
increases reservoir classification accuracies of smaller reservoirs,
the model still cannot work without perceptually training on its own
speech. Trying to extrapolate towards younger speakers with reservoirs
trained on older speakers proves more difficult than vice versa, as
we saw in section \ref{subsec:Comparing-reservoir-training}. Speaker
normalisation is still a problem for this model of speech acquisition.

\paragraph{Phonetic accuracy}

As mentioned in section~\ref{subsec:Vowel-shape-settings}, vowels
were labeled simply by me listening to them. Since this is subjective,
a next step would be to have them labeled by a phonetically trained
expert. However, I do not think that this will influence the learning
very much, except if the labeling were done more strictly (rejecting
more obscure speech sounds as belonging to the null class).

\paragraph{Sampling strictness}

Samples were hand-labeled after being listened to. Like this, even
speech sounds that weren't quite perfect, but fell into a specific
perceptual category (e.g. were heard as $/a/$) were admitted as samples
for the reservoir training. This, one could argue, is problematic
for two reasons:
\begin{enumerate}
\item Samples in formant space are no longer distinguishable and categories
overlap strongly. This may confuse the classifying reservoir, especially
when trying to extrapolate to younger speakers.
\item Infant directed speech in reality is pronounced deliberately, using
even more distinct pronounciation (for example larger vowel space).
\textit{Listen and Babble}, in a way, tries to form perceptual targets
while listening to (at best) adult directed speech.
\end{enumerate}
It would certainly be admittable to only train with samples that show
very clear phonetic recognisability. One might, for example, remove
all samples (or classify them as belonging to the $null$ class) that
have formants that differ too much from their corresponding prototype
formants. Thus, samples would become better separated in formant space
and phonetically more easily distinguishable.

A question might be: when samples are more distinct from each other,
does the auditory system interpolate better? In order to test this,
one could repeat (with the new, 'stricter' samples) reservoir training
paradigm number 4 (from section~\ref{subsec:Comparing-reservoir-training})
and compare error rates with those in Figure~\ref{fig:Error rates}.

\paragraph{Motherese}

Larger, and more distinct formant\textit{-trajectories} do seem to
make infant directed vowels easier to distinguish from each other
(especially $/o/$ and $/u/$ as seen in Figure~\ref{fig:German-vowels-spoken}).

By only looking at the one recorded motherese sample in section \ref{subsec:Motherese}
we can not make significant claims as to whether motherese acoustic
features move nearer to those of the infant, and thus help the infant
learner to generalise. However, we can see that trajectories are more
distinct. Larger (and thus clearer) trajectories in formant space
could aid in the generalisation by adding another characteristic,
by which an infant (or an ESN) can distinguish a specific vowel better,
independent of the precise location in formant space. In the sample,
we saw infant-directed $/u/$ and $/o/$ mainly distinguishable by
their trajectory, and not by their position in formant space.

Using vocal gestures with large pitch variation over time in \textit{Listen
and Babble} (while still using VocalTractLab for training data) does
not seem feasible. We should consider using data from human mother-infant
speech. KIDS (Konstanz prosodically annotated infant-directed speech
corpus)~\cite{KIDS Corpus} seems like an ideal source, since it
consists of German speech (spoken to infants by their mothers). Individual
syllables are already labeled and one only has to extract those phonemes
from which perceptual targets are formed (e.g. only vowels).

\paragraph{Understanding imitation more fully}

For my thesis, I (frustratingly) lacked the time to fully investigate
the second stage of the model (imitation). Murakami showed that, when
the auditory system only trains on vowel samples produced by the very
same vocal tract as that of the imitating agent, that agent is able
to completely learn those vowels and attain a near-perfect target
match. It is now important to understand how learning works when no
longer one perceptual target is formed, but rather a more nebulous
perceptual target over multiple speakers. I had time to run some first
tests using a reservoir that was trained with all speakers. Further
questions are:
\begin{itemize}
\item How does the RL algorithm compare with random babbling (sampling random
points in the hypercube until a point is found that yields a high
enough reward)?
\item On average, how many samples do we need for each vowel in order to
reach a certain reward level? (Maybe learn multiple targets in parallel
instead of intrinsic motivation for better comparison?)
\end{itemize}
Implementing variable reward thresholds might be a way to more evenly
learn targets. We might, for instance, learn all targets up to a low
threshold (e.g. 0.2 \textendash{} 0.3) and only then try to improve
utterance by raising target reward thresholds individually.

Also, the current target could be chosen on the go based on the increase
in reward (learning progress of that specific target). The RL-agent
could train where reward increases fastest, thus focusing on making
easier progress first.

\paragraph{Interdependence of articulator positions}

A point mentioned by Murakami in his thesis is the possibility to
introduce (biologically realistic) constraints as to what positions
each articulator can take. For now, the RL-agent can freely choose
each parameter, independently to the next, resulting (for instance)
in extreme tongue shapes. Making articulator parameters interdependent
could make learning simpler by reducing the dimensionality of the
problem.

I suggest restricting the curvature of the tongue (in the transverse
plane) to smaller values (e.g. each only moving between 0.4 and 0.6
in relative coordinates, 0.5 being the neutral flat position) as a
first easy step.

\paragraph{Comparing with other models}

The \textit{Listen and Babble} model of vowel aquisition in infants
comes with obvious advantages over some other models out there. 

Murakami describes some of these advantages in \cite{Paper Murakami}.
Here, I'd like to compare our approach in \textit{Listen and Babble}
with two models not yet mentioned in Murakami's discussion in his
thesis. Murakami provides an interesting comparison with the following
models: DIVA, Kröger, Warlaumont, Frier-Oudeyer (see pages 56-57 in
\cite{Thesis Murakami}).

The reason I mention the following models is this: they both emphasise
the role of the caregiver, which in \textit{Listen and Babble} is
limited to the forming of ambient speech for the perceptual training.
In our model the caregiver is not needed when imitating. This will
be the biggest difference between \textit{Listen and Babble} and the
following models:

Asada's group \cite{Asada1,Asada}, due to their research field (robotics),
recognise the \textit{correspondence problem}. For instance, in phonetics,
they modeled vowel acquisition using an artificial articulatory system
(with five degrees of freedom). Because this articulatory system is
structurally very different to that of a human, their speech robot
faces difficulties finding actions that correspond to human actions
with similar results. Training a simple speech robot to match a caregiver's
phonetic categories, they show that the robot solves the correspondence
in this way: caregivers' speech is not imitated directly (as in ``how
can I reproduce the sound which my caregiver made?''). Rather, the
caregiver imitates the robot's cooing, which in turn enables the robot
to learn more vowel-like articulations by mapping the phonetic regions
it can produce to those that the caregiver uses to imitate. Through
unconscious anchoring in the caregiver's imitation the robot bridges
the correspondence problem. The robot's phonetics approached those
of the caregiver without actually imitating the caregiver's speech.

The I.S. Howard and P. Messum model ELIJA \cite{Howard_1,Howard_2,Messum}
expands Asada's approach from vowel acquisition to learning syllables
and first words. They stress the fact that infants experience the
same correspondence problem (between their infantile vocal tracts
and those of the caregivers). Their model ELIJA learns equivalence
relations between its own vocal actions and the caregiver's speech
in response to those actions. ELIJA does not listen to the sound it
produced as our model does.

In my opinion, Howard and Messum make strong arguments for the importance
of caregiver imitation\footnote{Caregiver imitation: The caregiver responding to infant speech sounds
with imitated, phonetically more correct, adult speech sounds.} in solving the correspondence problem (which is similar to solving
the problem of speaker normalisation). But I do not know of ELIJA
accounting for spontaneous infant babbling without the constant presence
of a caregiver. Such babbling seems to be motivated by the actual
infant-produced sounds, which is a basic learning mechanism in \textit{Listen
and Babble}.

Might there be any way of combining the strengths of both types of
model (those based on how caregivers repond to infant actions and
those based on the infant directly imitating perceptual targets)?
The next and final section will attempt to answer that question.\cleardoublepage{}

\section{\label{CaregiverModel}Listen and Babble with Caregiver Imitation}

\bigskip{}

In this last section, I would like to propose an extension of \textit{Listen
and Babble}, which:
\begin{itemize}
\item no longer separates perception and imitation into two different stages
and
\item attempts to solve the perceptual problem of the infant having to associate
his own speech with that of his ambient speech (e.g. caregiver).
\end{itemize}
I suggest that the following model would:
\begin{itemize}
\item keep some of the advantages of \textit{Listen and Babble} over other
models of speech acquisition and
\item explain the role of a caregiver \textit{and} account for infants'
babbling in the absence of the caregiver.
\end{itemize}

\paragraph{A model including caregiver imitation\protect \\
}

As in \textit{Listen and Babble}, the infant's auditory system is
trained (perceptual stage). However, this is done without including
speech sounds produced by the infant vocal tract. This would be realistic:
babbling infants simply don't hear articulate speech from (other)
infants as young as themselves.

In \textit{Listen and Babble}, only the infant himself listens to
his speech sounds. I suggest that we include a caregiver who also
listenes to the infant's speech sounds. Caregiver imitation pairs
every adequate infant sound with an adult speech sound of that very
vowel. The infant would thus associate his own speech sounds, whenever
good enough to be imitated, with adult perceptual categories.

The auditory system is then retrained (after each generation of speech
samples), including those RL-agent's own sounds which were imitated
by the caregiver, as well as the adult imitation itself. This auditory
system returns confidences to the reinforcement learner. The infant
auditory system is basically retrained after each generation of samples.

A question that remains: \textit{how does the infant's auditory system
retrain?}

As in the sensory training in \textit{Listen and Babble}, a training
set is produced (with ambient speech samples). This training set first
consists of only non-infant speech samples (or e.g. samples from age
4 upward). Then, with every generation of self-produced speech, some
new speech sounds (infant sounds, each with a corresponding imitated
adult sound, and thus correctly labeled through association) are included
in the training set. In each generation, new sounds replace 'older'
sounds in the training set and in each generation, the auditory system
is trained anew with the training set.

Over time, the quality of the infant's auditory system will increase
for categorising its own sounds. I expect that the reinforcement learner
will also yield increasingly good results as a consequence.

We could easily disable the caretaker at any time (or at certain intervals),
thus returning to the original \textit{Listen and Babble} model. How
is learning influenced, when a caretaker is always present, constantly
prompting the auditory system to be retrained? Or what happens if,
after a while, the infant babbles along on his own for some time...?

\begin{figure}
\includegraphics[width=1\columnwidth]{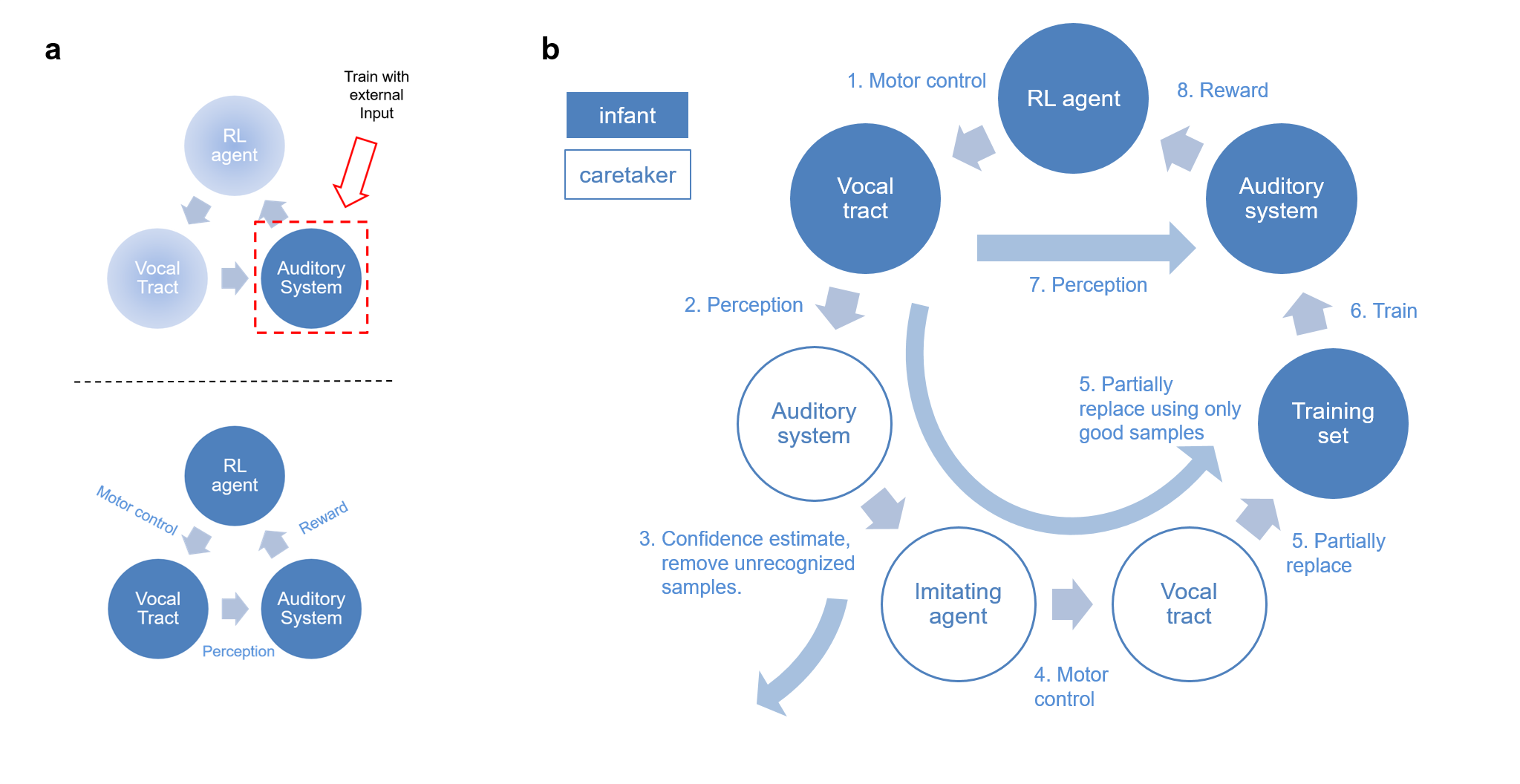}\caption{\label{fig:New_model}\textbf{a}: The \textit{Listen and Babble} model.
Two separate learning stages (target acquisition\textit{ top} and
imitation \textit{bottom})\protect \linebreak{}
\textbf{b}: A model including caregiver imitation. }
\end{figure}

In Figure~\ref{fig:New_model}, I schematically compare the original
\textit{Listen and Babble} model with the proposed extension. In step
five, infant speech samples will be correctly labeled due to the fact
that the caregiver has categorised them correctly. Providing an imitating
sound for each sample, these pairs are simply given the same label
before including them in the training set.

In Table \ref{tab:New_Algorithm}, I sketched a possible algorithm
for realising such a model. Some explanatory points ahead:
\begin{itemize}
\item The algorithm starts off with a fully trained caregiver auditory system
(trained on infant + adult speech of all ages), an infant auditory
system only trained on speech over a certain age. In a real sense
the infant learner (realistically) starts off without any prior knowledge
about:
\begin{itemize}
\item how his own speech should sound (no perceptual target),
\item how to produce speech (the same as in \textit{Listen and Babble}).
\end{itemize}
\item The algorithm needs to work in parallel with:
\begin{itemize}
\item $N_{G}$- Generation size (could be 1000, with 100 CPUs!),
\item $N_{I}$ \textendash{} A hand-set size of the 'best samples' in the
current generation. (For example: if $N_{G}=1000$ and $N_{I}=10$,
the best 1\% of the generation will be imitated by the caregiver).
\end{itemize}
\item The caregiver agent could even get confidences in respect to all the
classes, then label each sample with the class that yields maximum
confidence. The infant will then receive imitated speech sounds from
more then one class. This means, we could let the RL-Agent be intrinsically
motivated and swap the current target on the go.
\item Whenever the training set is ``updated with the current generation's
samples'', only samples produced in this imitation are replaced.
The training set has a fixed set of ambient speech (from, say, age
4 upwards). After all, the infant would continue to perceive ambient
speech. Only his own perceptual target (produced by the samples of
age 1 in the training set) shifts.\medskip{}
\end{itemize}
\begin{center}
\begin{table}
\noindent \begin{raggedleft}
\begin{tabular}{|>{\raggedright}p{0.05\columnwidth}|>{\centering}p{0.09\columnwidth}|>{\centering}p{0.5\columnwidth}|>{\centering}p{0.3\columnwidth}|}
\hline 
{\scriptsize{}Who?} & \raggedright{}{\scriptsize{}Com-}\\
{\scriptsize{}ponent} & \raggedright{}\textcolor{black}{\scriptsize{}Tasks} & \raggedleft{}\textit{\scriptsize{}Resulting data (Size)}\tabularnewline
\hline 
\hline 
\multirow{2}{0.05\columnwidth}{{\scriptsize{}Infant}} & \raggedright{}{\scriptsize{}RL-}\\
{\scriptsize{}Agent} & \begin{raggedright}
\textcolor{black}{\tiny{}- choses x\_mean (from cma-es)}
\par\end{raggedright}{\tiny \par}
\begin{raggedright}
\textcolor{black}{\tiny{}- generates $N_{G}$ offspring}
\par\end{raggedright}{\tiny \par}
\raggedright{}\textcolor{black}{\tiny{}- return infant\_parameters} & \raggedleft{}\textit{\scriptsize{}infant\_parameters($N_{G}$)}\tabularnewline
\cline{2-4} 
 & \raggedright{}{\scriptsize{}Vocal }\\
{\scriptsize{}tract} & \begin{raggedright}
\textcolor{black}{\tiny{}- synthesise infant speech using infant\_parameters}
\par\end{raggedright}{\tiny \par}
\raggedright{}\textcolor{black}{\tiny{}- return infant\_speech\_sounds
({*})} & \raggedleft{}\textit{\scriptsize{}infant\_parameters($N_{G}$),}\\
\textit{\scriptsize{} infant\_speech\_sounds($N_{G}$)}\tabularnewline
\hline 
\multirow{3}{0.05\columnwidth}{{\scriptsize{}Care-}\\
{\scriptsize{}giver}} & \raggedright{}{\scriptsize{}Auditory}\\
{\scriptsize{}System} & \begin{raggedright}
\textcolor{black}{\tiny{}- perception of infant\_speech\_sounds}
\par\end{raggedright}{\tiny \par}
\raggedright{}\textcolor{black}{\tiny{}- return caregiver confidences
({*}{*})} & \raggedleft{}\textit{\scriptsize{}infant\_parameters($N_{G}$),}\\
\textit{\scriptsize{} infant\_speech\_sounds($N_{G}$),}\\
\textit{\scriptsize{}cg\_confidences($N_{G}$)}\tabularnewline
\cline{2-4} 
 & \raggedright{}{\scriptsize{}Cgv.-}\\
{\scriptsize{}Agent} & \begin{raggedright}
\textcolor{black}{\tiny{}- sort data according to each sample's maximal
confidence value ({*}{*}{*})}
\par\end{raggedright}{\tiny \par}
\begin{raggedright}
\textcolor{black}{\tiny{}- keep first $N_{I}$ of samples only, reject
all others}
\par\end{raggedright}{\tiny \par}
\begin{raggedright}
\textcolor{black}{\tiny{}- imitate all $N_{I}$ samples with their
nearest class teacher parameters. These will be adult parameters.}
\par\end{raggedright}{\tiny \par}
\raggedright{}\textcolor{black}{\tiny{}- return adult\_parameters
($N_{I}$ ) for each of those $N_{I}$ infant samples.} & \begin{raggedleft}
\textit{\scriptsize{}infant\_parameters($N_{I}$),}\\
\textit{\scriptsize{} infant\_speech\_sounds($N_{I}$), cg\_confidences($N_{I}$),}\\
\textit{\scriptsize{} adult\_parameters($N_{I}$),}
\par\end{raggedleft}{\scriptsize \par}
\raggedleft{}\textit{\scriptsize{}adult\_parameters($N_{I}$)}\tabularnewline
\cline{2-4} 
 & \raggedright{}{\scriptsize{}Vocal }\\
{\scriptsize{}tract} & \begin{raggedright}
\textcolor{black}{\tiny{}- synthesise adult speech using adult\_parameters}
\par\end{raggedright}{\tiny \par}
\raggedright{}\textcolor{black}{\tiny{}- return adult\_speech\_sounds
($N_{I}$ )} & \begin{raggedleft}
\textit{\scriptsize{}infant\_parameters($N_{I}$), infant\_speech\_sounds($N_{I}$),
cg\_confidences($N_{I}$), adult\_parameters($N_{I}$),}
\par\end{raggedleft}{\scriptsize \par}
\begin{raggedleft}
\textit{\scriptsize{}adult\_parameters($N_{I}$),}
\par\end{raggedleft}{\scriptsize \par}
\raggedleft{}\textit{\scriptsize{}adult\_speech\_sounds($N_{I}$)}\tabularnewline
\hline 
\multirow{2}{0.05\columnwidth}{{\scriptsize{}Infant}} & \raggedright{}{\scriptsize{}Auditory }\\
{\scriptsize{}System} & \begin{raggedright}
\textcolor{black}{\tiny{}- perceptual labeling:}
\par\end{raggedright}{\tiny \par}
\begin{raggedright}
\textcolor{black}{\tiny{}Each adult speech sound comes with the corresp.
infant speech sound (pairs). Perception of each adult sound yields
a label for the corresp. infant sound.}
\par\end{raggedright}{\tiny \par}
\begin{raggedright}
\textcolor{black}{\tiny{}- replace 'oldest'({*}{*}{*}{*}{*}) samples
in the TRAINING\_SET ({*}{*}{*}{*}) with the new labeled infant samples..}
\par\end{raggedright}{\tiny \par}
\begin{raggedright}
\textcolor{black}{\tiny{}- retrain reservoir with updated TRAINING\_SET
(which has now been updated with the 'good' samples)}
\par\end{raggedright}{\tiny \par}
\begin{raggedright}
\textcolor{black}{\tiny{}- (improved?) perception of infant\_speech\_sounds}
\par\end{raggedright}{\tiny \par}
\raggedright{}\textcolor{black}{\tiny{}- return inf\_confidences} & \begin{raggedleft}
\textit{\scriptsize{}inf\_confidences($N_{I}$)}
\par\end{raggedleft}{\scriptsize \par}
\raggedleft{}{\scriptsize{}(}\textit{\scriptsize{}Reset others)}\tabularnewline
\cline{2-4} 
 & \raggedright{}{\scriptsize{}RL-Agent} & \begin{raggedright}
\textcolor{black}{\tiny{}- compute reward of best sample}
\par\end{raggedright}{\tiny \par}
\raggedright{}\textcolor{black}{\tiny{}- return to beginning (new
x\_mean)} & \raggedleft{}\tabularnewline
\hline 
\end{tabular}
\par\end{raggedleft}
\noindent \caption{\label{tab:New_Algorithm}A sketch of a possible algorithm to implement
caregiver imitation.\protect \\
{\scriptsize{}({*})}\textit{\scriptsize{} infant\_speech\_sounds}{\scriptsize{}
could be a list of paths to the wav files.}\protect \\
{\scriptsize{}({*}{*}) each sample will have a list of confidences
for each class of the Auditory System}\protect \\
{\scriptsize{}({*}{*}{*}) the maximum is taken over the classes.}\protect \\
{\scriptsize{}({*}{*}{*}{*}) TRAINING\_SET: initially consists of
labeled samples from speakers over a certain age (e.g. 4~yrs). This
global set of labeled speech sounds is regularly updated in each iteration
and repeatedly used to train another reservoir for the infant's auditory
system.}\protect \\
{\scriptsize{}({*}{*}{*}{*}{*}) }\textcolor{black}{\scriptsize{}Replace
only those samples produced by the infant learner (not ambient speech).
Initially, no samples are replaced, however after mult. generations,
older infant speech in the training set must be replaced in order
to make progress.}}
\end{table}
\par\end{center}

\noindent An objection my readers might have:

\noindent \textit{An auditory system that is (even at first) only
trained on bad samples (those of maximal confidence will still not
be good) cannot return reward signals accurate enough for the reinforcement
learner to make any progress!}

That, I believe, is true if the auditory system is \textit{only }trained
on those 'bad samples'. But in fact, the (large portion) of training
data (that of all older speakers) will be accurate. The additional
infant's samples are only there to 'help bridge the gap' between its
own sounds and those of, say, 4 year olds and upward. Having some
(non-perfect) speech sounds of the infant himself included in the
perceptual retraining might help the infant to generalise across speakers.

\cleardoublepage{}

\section{\label{sec:For-developers}For developers}

\subsection{Getting started}

{\footnotesize{}Each step in 'Results' is also documented in the project
source code and can be reproduced by:}{\footnotesize \par}
\begin{enumerate}
\item {\footnotesize{}downloading }\textit{\footnotesize{}Listen and Babble}{\footnotesize{}
from~\cite{Repository},}{\footnotesize \par}
\item {\footnotesize{}installing needed dependencies,}{\footnotesize \par}
\item {\footnotesize{}executing the shell while having the relevant lines}\footnote{{\footnotesize{}Example:}\\
{\footnotesize{}In }\textit{\footnotesize{}control/get\_params.py:}{\footnotesize \par}

\textcolor{gray}{\footnotesize{}self.execute\_main\_script = \{'ambient\_speech':True,'hear':False,'learn':False\}}{\footnotesize \par}

\textcolor{gray}{\footnotesize{}self.do\_setup = True }{\footnotesize \par}

\textcolor{gray}{\footnotesize{}self.do\_make\_proto = True}{\footnotesize \par}

\textcolor{gray}{\footnotesize{}self.do\_setup\_analysis = True}\\
{\footnotesize{}Then check results in results/ambient\_speech/srangefm(2)/..}{\footnotesize \par}}{\footnotesize{} in }\textit{\footnotesize{}control/get\_params.py}{\footnotesize{}
set as $True$,}{\footnotesize \par}
\item {\footnotesize{}checking result plots in the result directory or 'pickle-loading'
the saved class in }\textit{\footnotesize{}data/classes/.. .pickle.}{\footnotesize \par}
\end{enumerate}

\subsection{Project structure}

\textit{Listen and Babble} is implemented using Python. I structured
the code using \textit{classes}, in which I grouped all functions
belonging to one main stage of the project. Each stage of the project
is executed from a project shell (in the main directory). The shell
instantiates main classes, which in turn call functions from the functions
class.

\begin{figure}
\includegraphics[width=1\columnwidth]{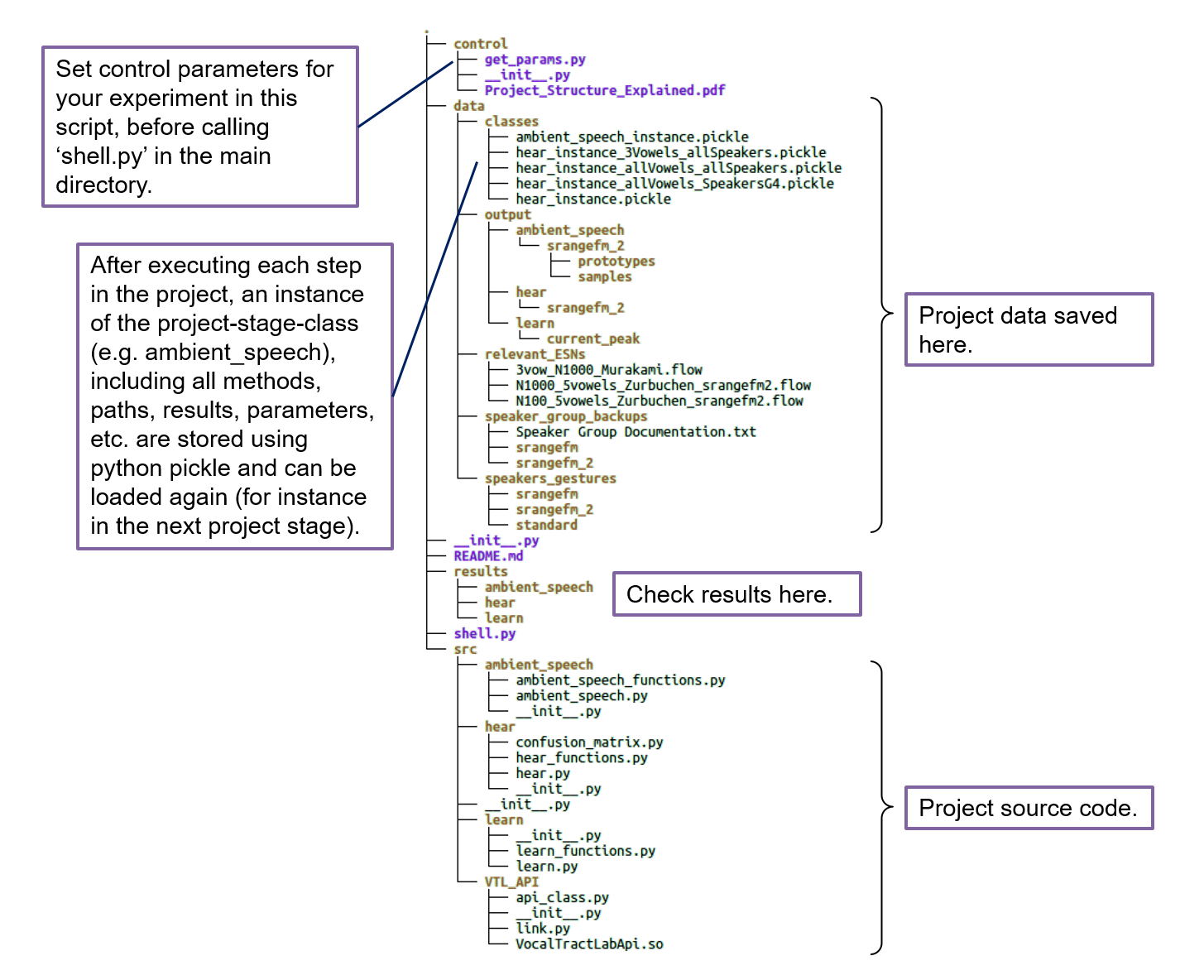}

\caption{\label{fig:Project structure}Project structure of my implementation
of \textit{Listen and Babble} \cite{Repository}.}

\end{figure}

For an overview of the project directory, see Figure~\ref{fig:Project structure}.

I advise first reading the documentation in the main directory{\footnotesize{}~\cite{Repository}},
then shell and get\_params scripts, and after that to venture on to
understand specific functions.

\noindent 

\cleardoublepage{}

\appendix

\part*{Appendix\cleardoublepage{}}

\title{\includegraphics[width=1\columnwidth]{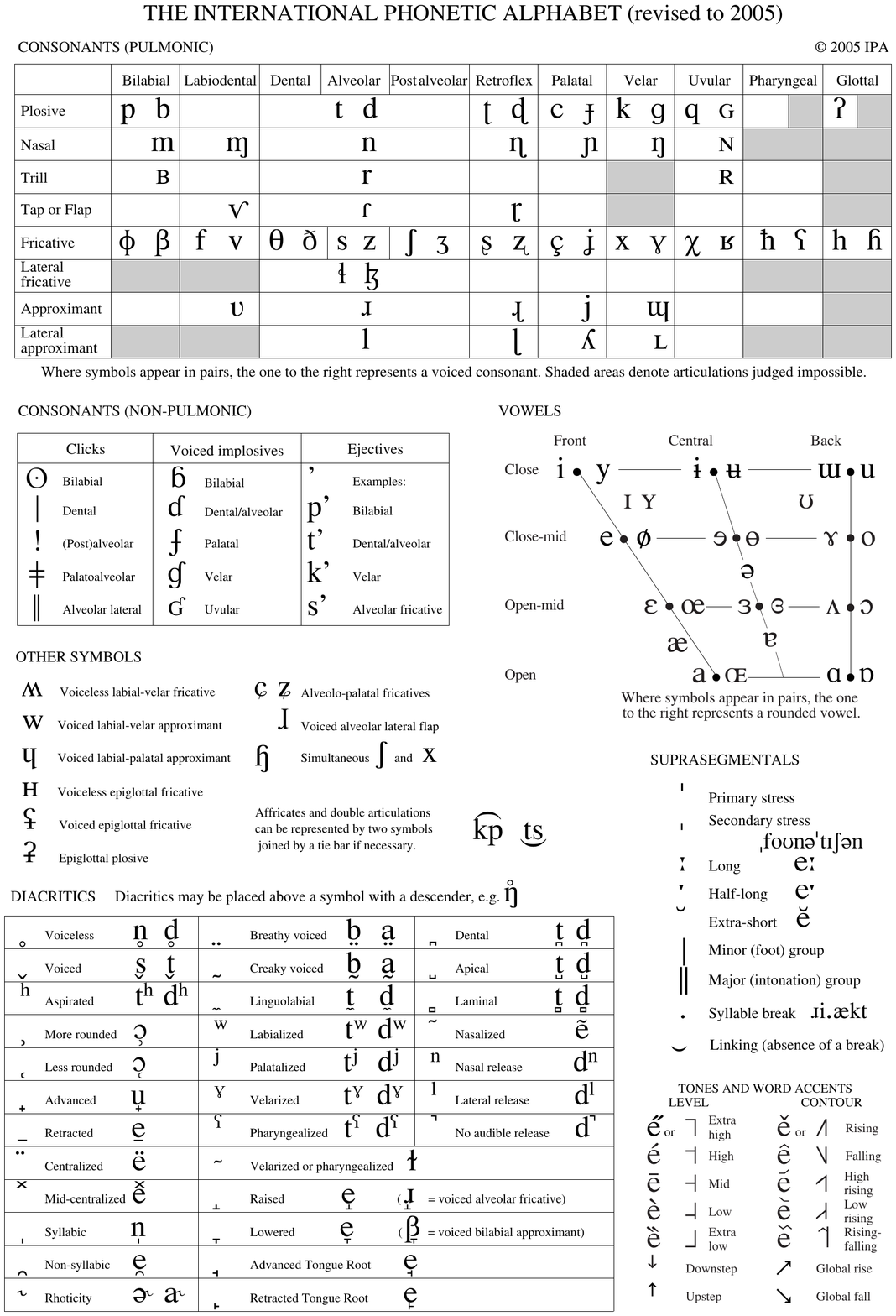}\cleardoublepage{}}

\bigskip{}
Erklärung nach $\mathsection$ 30 (12) Ordnung für den Bachelor- und
den Masterstudiengang:

\medskip{}

Hiermit erkläre ich, dass ich die Arbeit selbstständig und ohne Benutzung
anderer als der angegebenen Quellen und Hilfsmittel verfasst habe.
Alle Stellen der Arbeit, die wörtlich oder sinngemäß aus Veröffentlichungen
oder aus anderen fremden Texten entnommen wurden, sind von mir als
solche kenntlich gemacht worden. Ferner erkläre ich, dass die Arbeit
nicht \textendash{} auch nicht auszugsweise \textendash{} für eine
andere Prüfung verwendet wurde.

\bigskip{}

Frankfurt, den\\

\end{document}